\documentstyle[prb,aps,epsf,eqsecnum,twocolumn]{revtex}

\begin{document}
\draft

\twocolumn[\hsize\textwidth\columnwidth\hsize\csname@twocolumnfalse\endcsname
\title{Fractional Excitations in the Luttinger Liquid.}
\author{K.-V. Pham, M. Gabay and P. Lederer}
\address{Laboratoire de Physique des Solides, associ\'e au CNRS \\
Universit\'{e} Paris--Sud \\
91405 Orsay, France}
\date{\today}
\maketitle
\begin{abstract}
{
We reconsider the spectrum of the Luttinger liquid (LL) usually understood in terms of phonons (density fluctuations),
 and within the context of bosonization we 
give an alternative representation in terms of fractional states. This allows to make contact with
Bethe Ansatz which predicts similar fractional states. As an example we study the spinon operator in the absence of spin rotational
invariance and derive it from first principles: we find that it is not a
semion in general; a trial Jastrow wavefunction is
also given for that spinon state. Our construction of the new spectroscopy based on fractional states 
leads to several new physical insights: in the low-energy limit, we find that the $S_{z}=0$ continuum of gapless spin
chains  is due to pairs of
fractional quasiparticle-quasihole states which are the 1D counterpart of the Laughlin FQHE
quasiparticles. The holon operator for the Luttinger liquid with spin is
also derived. In the presence of a magnetic field, spin-charge separation is
not realized any longer in a LL: the holon and the spinon are then replaced
by new fractional states which we are able to describe.

}
\end{abstract}
] \widetext
\narrowtext

\section{Introduction.}

\label{section-un}

\subsection{Motivations of this work.}

\label{sub1-1}

One of the most striking property of some strongly correlated systems is
fractionalization, that is the existence of elementary excitations carrying
only part of the quantum numbers of the constituent particles of the system.
The most famous example is probably the charge one-third Laughlin
quasiparticle, which is the elementary excitation of the fractional quantum
Hall fluid at filling $\nu =1/3$.\cite{laughlin} Its existence was recently
confirmed in a beautiful set of shot noise experiments\cite{glattli}. The
earliest example of fractionalization in condensed matter physics is however
found in one dimension: the exact solution of the Hubbard model\cite{hubbard}
by Bethe Ansatz\cite{bethe} revealed that the charge and spin of the
electron split into two excitations with independent dynamics, known as the
holon and the spinon. Faddeev and Takhtajan later showed that the same
spinon is also the elementary excitation of the 1D Heisenberg model: the
magnon (the usual Goldstone boson) is replaced by two spinons generating a
continuum for $\Delta S=1$ excitations\cite{fadeev,johnson}. This property
of the Hubbard model is known as spin-charge separation and is generic of
so-called Luttinger liquids (LL): LL constitute a universality class for
gapless one dimensional models such as the Heisenberg chain, the Hubbard and 
$t-J$ models\cite{haldane79}. Luttinger liquids are non-Fermi liquids: Landau quasiparticles%
\cite{landau} are not elementary excitations of the LL and as a consequence
the electron Green's function shows no quasiparticle pole (this property is
true both for the LL with spin and for the spinless LL). Haldane, who coined
the name of LL, conjectured that 1D gapless models would have the same
low-energy physics as that of the Tomonaga-Luttinger model. For
energies smaller than the bandwidth\cite
{luttinger,tomonaga,mattis-lieb}, the latter model is
a fixed point of the renormalization group (RG)\cite{shankar}. In 1D, bosonization allows to
transform the Tomonaga-Luttinger model into a gaussian acoustic hamiltonian
describing free phonons \cite
{f1}; the considerable success and popularity of bosonization stems from the
fact that all the computations are straightforward because the effective
hamiltonian is that of a free bosonic field. Another perspective on the LL is provided by conformal field theory (CFT) which describes
two dimensional (or $1+1$) critical theories with conformal invariance; this
has allowed to identify the Luttinger liquid universality class as the set
of $c=1$ CFTs\cite{anomaly}, i.e. the set of all models which flow under RG
towards the gaussian free boson hamiltonian\cite{cftref}. CFT has allowed to
formalize the finite-size analysis of Luttinger liquids first introduced by
Haldane\cite{cardy,kawakami,haldane79}. In terms of the gaussian
hamiltonian, the LL theory can be described as a phenomenological theory
characterized by the following parameters: $u$ which is a velocity for
collective modes and $K$ which is proportional to the compressibility of the
system.

Yet, although the LL description is supposedly quite well established through
the formalisms of bosonization or CFT, and despite the fact that exact solutions
(Bethe Ansatz)  show the existence of fractional states in the spectrum
of several Luttinger liquids, there exists no systematic study of fractional
excitations in the LL to the best of the authors' knowledge. What's more,
in the framework of the bosonization formalism, it is sometimes stated that the
only physically relevant excitations of a LL are phonons, since the
effective hamiltonian is just that of acoustic phonons. As we show below
this statement is incorrect. Conformal field
theory is an alternative to bosonization which does stress the spectroscopic
aspects: yet, application to the study of fractional excitations in a LL has
been limited to the spinon in the case of $SU(2)$ symmetry, which is the
situation relevant for the Heisenberg chain\cite{schoutens}. Fractional
excitations must exist in a Luttinger liquid if Bethe Ansatz is correct but
as far as the authors are aware, the characterization of these very
unconventional fractional states, through either bosonization or CFT, is
mostly {\it terra incognita} as the following list of issues may show:

1) $\Delta S=1$ excitations for the Heisenberg chain form a continuum of
pairs of spinons. When an Ising anisotropy is introduced, in the massless
regime (with obvious notations: $\left| J_{z}\right| \leq J_{x}$ $(=J_{y})$
) the continuum still exists and evolves smoothly as a function of the
anisotropy $\Delta =J_{z}/J_{x}$\cite{johnson}. The continuum is again
ascribed to pairs of spinons\cite{spinon}. It is intuitively clear that
these spinons should be in some sense deformations of the $SU(2)$ spinon. We
will derive in this paper creation operators for these non-$SU(2)$ spinons
using the bosonization method.

In the case of $SU(2)$ symmetry, trial wavefunctions for these spinons can
be found by making use of the exact solvability of the Haldane-Shastry (HS)
chain\cite{hs}. The HS model shares the properties of the Heisenberg chain:
it is a gapless $SU(2)$ symmetric spin chain with a continuum of spinon
excitations. Its ground state and spinon wavefunctions\cite{hal94} are
remarkably similar to those one would write for bosonic Laughlin states at
filling $\nu =1/2$:

\begin{eqnarray}
\Psi _{gs}(x_{1},..,x_{N}) &=&\prod_{i<j}(z_{i}-z_{j})^{2}, \\
\Psi _{spinon}(z_{0}) &=&\prod_{i}(z_{i}-z_{0})\prod_{i<j}(z_{i}-z_{j})^{2},
\\
z_{i} &=&\exp i\frac{2\pi }{L}x_{i},
\end{eqnarray}
where $x_{i}$ is the coordinate a spin down, and $x_{0}$ that of the spinon%
\cite{phase}. We will exhibit a similar wavefunctions for the
spinon in the absence of $SU(2)$ symmetry. $SU(2)$ spinons are semions\cite
{hal94} (anyons with a statistics intermediate between that of fermions and
bosons): we will show that the statistics is affected when an anisotropy is
introduced.

A continuum is also found by Bethe Ansatz for $\Delta S_{z}=0$ transitions 
\cite{woy82}. Low-lying excitations are described in that approach as
two-strings states in the string formalism customary to Bethe Ansatz. This
description is similar to that given for the $\Delta S=1$ continuum of the
isotropic chain. In the latter case it is quite clear that a spin one-half
should be ascribed to each of the (pseudo) ''hole'' states in the string,
which leads to the spinon interpretation since each state should contribute
symmetrically to the spin-flip\cite{fadeev}. For $\Delta S_{z}=0$
transitions, the total $z$ spin components of the excitations add up to $0$
and the continuum results from the excitation of particle-hole pairs. For the isotropic chain, this continuum is generated by spin $1/2$ 
spinon-antispinon pairs. 
By contrast, for the $XY$ chain the $\Delta S_{z}=0$ continuum is
due to particle-hole pairs of magnon-like spin $S_{z}=1$ excitations.
 The case of the isotropic Heisenberg chain
for which spin $1/2$ spinons are involved both in the $\Delta S_{z}=0$ and
the $\Delta S_{z}=1$ continuum, is therefore incidental. The important
lesson to be learned is that in the presence of an Ising anisotropy, the $%
\Delta S_{z}=1$ and $\Delta S_{z}=0$ continua may involve different
fractional spin states: in the first case we have spinons\cite{spinon}, but
in the second case the spinon identification is not always correct. What
happens in the case of an arbitrary anisotropy will be dealt with in this
paper.

2) The holon appearing in the exact solution of the Hubbard model is a
spinless charge one excitation\cite{lieb}. The issues raised for the spinon
(operator, wavefunction, statistics) extend naturally to the holon.

3) Spin-charge separation is an asymptotic property of the Hubbard model
valid in the low-energy limit. When a magnetic field is applied Frahm and
Korepin found that spin-charge decoupling was not realized even in the
low-energy limit\cite{frahm}. In this paper we derive the new excitations replacing the holon and the spinon.

4) In the context of the Calogero-Sutherland (CS) model\cite{cs} the
existence of fractional excitations similar to Laughlin quasiparticles was
suggested\cite{zirn}. In a variant of the standard LL known as the chiral LL
used to describe edges of a FQHE sample, Laughlin quasiparticles do appear
but the existence of such states follows from that of the same excitations
in the bulk\cite{wen}. In the CS model the proposal was triggered by the
similarity of the ground state with that of the 2D Laughlin wavefunctions
and by special selection rules. The ground state is\cite{cs}:

\begin{eqnarray}
\Psi (x_{1},..,x_{N}) &=&\prod_{i<j}\left| z_{i}-z_{j}\right| ^{\lambda }, \\
z_{i} &=&\exp i\frac{2\pi }{L}x_{i}
\end{eqnarray}
where $\lambda $ is a coupling constant for the $1/r^{2}$ interaction
potential of the CS model. The CS model is a LL\cite{kawakami} and the LL
parameter is just $K=1/\lambda $. A pseudo-particle formalism similar to
that of Bethe Ansatz can be introduced and for the restricted case of
rational couplings $\lambda =p/q$ special selection rules are found for the
dynamical structure factor: $p$ pseudo-holes (a pseudo-hole is a hole in the
Fermi sea of pseudo-momenta) must be accompanied in any excitation by $q$
pseudo-particles\cite{ha}. For a charge $(-1)$ pseudo-particle, this means
one has a charge $1/\lambda =K$ for the pseudo-hole. In the interpretation
of those selection rules it is proposed to view the CS model as a gas of
non-interacting pseudo-particles with anyonic statistics $\pi \lambda $ and
one rewrites the ground state as an anyonic wavefunction $\prod_{i<j}\left(
z_{i}-z_{j}\right) ^{\lambda }$. The pseudo-holes are particle-hole
conjugate of these anyons: the main modification with the non-interacting
case being the new selection rule; a wavefunction for these pseudo-holes
consistent with those interpretations is then $\prod_{i}\left(
z_{i}-z\right) \prod_{i<j}\left( z_{i}-z_{j}\right) ^{\lambda }$ which has
the correct charge and statistics. The pseudo-hole is therefore identified
as a Laughlin quasiparticle. It exists for rational couplings and carries
the rational charge $1/\lambda $.

There are however several limitations to those views. Firstly, these
considerations are only valid for rational couplings (the pseudo-particle
selection rules can not be extended to irrational $\lambda $): the physics
of the CS model is by contrast completely continuous with the coupling and
does not discriminate between rational and irrational couplings. The
impossibility to describe rational and irrational couplings on the same
footing means the representation is not adequate. Secondly a disymmetry
between particle and hole excitations is introduced.

In a parallel strand of ideas, Laughlin quasiparticles were also proposed in
studies of transport in a LL. The basis of the argument is that for a LL
with an impurity potential a charge $K$ and not a charge unity is
backscattered at the impurity location (where $K$ is the LL parameter, i.e.
the conductance of the LL)\cite{fisher}. The impurity potential can be
rewritten as a hopping potential for a charge $K$ state whose exchange
statistics is $\pi K$ as seen from the commutation relations (i.e. anyonic).
But for $K=1/(2n+1)$ these states are just those given by Wen for the
Laughlin quasiparticles of his chiral LL: this suggests to identify these
states as Laughlin quasiparticles. The main difficulty in that argument is
that it relies on the introduction of an impurity potential in the LL: this
obscures the question of the existence or not of a Laughlin quasiparticle in
the pure non-chiral Luttinger liquid. In summary what is missing is a proof
that states similar to Laughlin quasiparticles might be exact eigenstates of
the LL boson hamiltonian (i.e. of the RG fixed point in the low-energy
limit). The existence of Laughlin quasiparticles for the non-chiral LL for
arbitrary couplings must then be considered at this point as an unproved
conjecture.

The long list of issues we have brought up in points $(1-4)$ above should
convince the reader that a thorough discussion of fractional excitations for
Luttinger liquids within the formalisms of bosonization or CFT remains to be
done. This is what motivated us to re-examine in that paper the spectrum of
Luttinger liquids. We want to stress that although the previous examples
concern integrable models, the detailed physics of such integrable
models is not really our main interest: what matters for us is the universal
low-energy content of these theories and of course we will be unable to tell
anything through bosonization on the high-energy physics. Although it seems to be taken for granted that the excitations of a LL are the holon and the
spinon on account of Bethe Ansatz studies of the Hubbard model, we are not
aware of any existence proof of such fractional excitations whenever the
model is non-integrable: this is so because any proof must resort to the
universality hypothesis, that is to the LL and bosonization frameworks. The
theoretical formalism we wish to introduce aims at bridging that gap by
focusing on the universal structure of fractional excitations of Luttinger
liquids through the bosonization method.

The structure of the paper will be as follows: section \ref{section-un} is
an introduction to the topics considered in the paper. In subsection \ref
{sub1-2} we give a short review of the LL physics in order to set the
notations used throughout the paper; the issues discussed in the present
subsection \ref{sub1-1} will be amplified in subsection \ref{sub1-23} in
which we present the standard view on excitations of the LL. In the next
section \ref{ch-two} we will show that an alternative eigenstate basis can
be built: that quasiparticle basis allows a natural discussion of fractional
states. In section \ref{ch-3}, we will generalize our analysis to the LL
with spin. When a magnetic field is added on to the Hubbard model, spin-charge
separation no longer occurs\cite{frahm}. The standard spin-charge separated
Luttinger liquid theory is not applicable any more. We will introduce in
subsection \ref{ch-32} a general framework related to the $K$ topological
matrix of Wen's chiral Luttinger liquids\cite{wen}, which yields simple
criteria of spin-charge separation in terms of a $Z_{2}$ symmetry: we will
be able then in subsection \ref{ch-33} to derive the fractional excitations
which replace the holon and the spinon. The general LL theory we have
introduced will then be applied to the Hubbard model in a magnetic field in
sub-section \ref{ch-34} in which we explain the relation between our
approach and the formalism of the dressed charge matrix due to Frahm and
Korepin. Let us mention that section \ref{ch-two} of this paper expands 
on a short version which
contained results in the case of a spinless LL\cite{short}, whereas part \ref{ch-3}
presents totally new material.

\subsection{The Luttinger liquid.}

\label{sub1-2}

\subsubsection{Notations.}

This section will define the notations employed throughout the paper. We
exclusively deal with Luttinger liquids and therefore when considering some
specific models such as the Heisenberg spin chain or the Hubbard model we
implicitly assume that we are working in the LL part of their phase
diagrams. The whole physics of the LL is embodied in the following
hamiltonian: 
\begin{equation}
H_{B}=\frac{u}{2}\int_{0}^{L}dx\ K^{-1}(\nabla \Phi (x))^{2}+K(\nabla \Theta
(x))^{2}  \label{boson}
\end{equation}
supplemented by the so-called bosonization formulas. We work on a ring of
length $L$. $u$ and $K$ are the LL parameters. $\Phi $ can be interpreted as
a displacement field for phonons, while $\Theta $ is a superfluid phase;
indeed the particle and current densities are defined as:

\begin{eqnarray}
\rho (x)-\rho _{0} &=&-\frac{1}{\sqrt{\pi }}\nabla \Phi (x), \\
j(x) &=&\frac{1}{\sqrt{\pi }}\nabla \Theta (x).
\end{eqnarray}
{\it Renormalized current:} Actually $j(x)$ is a {\it bare current density}
which corresponds to the correct one only in the non-interacting case $K=1$:
the continuity equation shows that the correct current density is
renormalized and is 
\begin{equation}
j_{R}(x)=uKj(x)
\end{equation}
(the Fermi velocity has been set to unity). We will discuss in
section \ref{ch-two} the meaning of such a renormalization.

The particle operators for bosons and right and left moving fermions
respectively are given as:

\begin{eqnarray}
\Psi _{B}(x) &=&:\exp i\sqrt{\pi }\Theta (x) :, \\
\Psi _{F,R}(x) &=&:\exp i\sqrt{\pi }\left( \Theta (x) -\Phi (x)\right) :\exp
ik_{F}x, \\
\Psi _{F,L}(x) &=&:\exp i\sqrt{\pi }\left( \Theta (x)+\Phi (x) \right) :\exp
-ik_{F}x
\end{eqnarray}
(in the following we will assume that these operators are normal ordered). $%
k_{F}=\pi N_{0}/L$ is the Fermi momentum where $N_{0}$ is the number of
particles which is fixed by the chemical potential. $\Theta $ and $\Pi
=\nabla \Phi $ are canonical conjugate boson fields: 
\begin{equation}
\left[ \Theta (x),\Pi (y)\right] =i\delta (x-y).
\end{equation}

The zero modes of the charge and current density are respectively:

\begin{eqnarray}
\widehat{N} &=&N_{0}+\widehat{Q}=\int_{0}^{L}\rho (x)dx=N_{0}-\int_{0}^{L}%
\frac{1}{\sqrt{\pi }}\nabla \Phi dx, \\
\widehat{J} &=&\int_{0}^{L}j(x)dx=\int_{0}^{L}\frac{1}{\sqrt{\pi }}\nabla
\Theta (x)dx.
\end{eqnarray}
$\widehat{Q}$ has integral eigenvalues as befits a charge operator; in the
bosonization mapping, the charge quantization is taken into account by the
topological quantization of the phase field $\Phi $. Similarly, since $%
\int_{0}^{L}j(x)dx$ is a closed line integral (around the LL), it is a
quantized number: this is just the topological quantization of the
superfluid phase; the normalization of the fields have been chosen so that $%
\widehat{J}$ is an integer. For fermions, $\widehat{Q}=N_{+}+N_{-}$ and $%
\widehat{J}=N_{+}-N_{-}$ where $N_{+}$ and $N_{-}$ are respectively the
(integral) number of (bare) electrons added to the ground state at the right
and left Fermi points. The construction we have reviewed above is due to
Haldane\cite{hal81}.

Integrating the Fourier expansions of the charge and current density gives:
\begin{eqnarray}
\Theta (x) &=&\Theta _{0}+\frac{\sqrt{\pi }}{L}\widehat{J}x+\frac{1}{\sqrt{L}%
}\sum_{n\neq 0}\Theta _{n}\exp i\frac{2\pi n}{L}x,  \label{theta} \\
\Phi (x) &=&\Phi _{0}-\frac{\sqrt{\pi }}{L}\widehat{Q}x+\frac{1}{\sqrt{L}}%
\sum_{n\neq 0}\Phi _{n}\exp i\frac{2\pi n}{L}x.  \label{phi}
\end{eqnarray}
\ Note that these fields are not periodic: this allows for the above
mentioned topological excitations. We demand that the boson or fermion
operators are physical objects and be periodic on the ring: $\Psi
_{B/F}(x)=\Psi _{B/F}(x+L)$; this then implies the following selection rules
on the eigenvalues $Q$ and $J$ of the zero modes $\widehat{Q}$ and $\widehat{%
J}$:

\begin{eqnarray}
Bosons &:&\;J\;even\;%
\mathop{\rm integer}%
, \\
Fermions &:&\;Q-J\;\;even\;%
\mathop{\rm integer}%
.
\end{eqnarray}
Both $Q$ and $J$ are integers. 
The zero modes are sometimes extracted from the definition of the fermion
operator which defines the $U_{\pm }$ operators, first built by Heidenreich
and Haldane for the Tomonaga-Luttinger model\cite{heid,haldane79}: 
\begin{equation}
U_{\pm }=\exp i\sqrt{\pi }\left( \Theta _{0}\pm \Phi _{0}\right) .
\end{equation}
It will be useful to consider the commutation properties of the following
operators: 
\begin{equation}
V_{\alpha ,\beta }(x)=:\exp -i\sqrt{\pi }\left( \alpha \Theta (x)-\beta \Phi
(x)\right) :.
\end{equation}
Using Campbell-Haussdorf formula, one finds:

\begin{eqnarray}
&&V_{\alpha ,\beta }(x)V_{\alpha ,\beta }(y)  \nonumber \\
&=&V_{\alpha ,\beta }(y)V_{\alpha ,\beta }(x)e^{-i\pi \alpha \beta
\;sgn(y-x)}  \label{anyon}
\end{eqnarray}
where $sgn(x)$ is the sign function,
which shows in particular that $\Psi _{F}(x)$ is a fermionic
operator. We define the exchange statistics of an operator per: 
\begin{eqnarray}
&&O(x)O(y)  \nonumber \\
&=&O(y)O(x)\exp -i\theta \;sgn(y-x).
\end{eqnarray}
For instance $\theta =\pi $ for fermions.

\subsubsection{Excitations.}

\label{sub1-23}

Until the work of Heidenreich\cite{heid} and subsequently { of} Haldane\cite
{haldane79}, the only excitations considered in the gaussian model were the
bosonic phonon (or plasmon\cite{f1}) modes. But the hamiltonian contains a
second part corresponding to the energies of states with
non-zero charge or current with respect to the ground state. In reciprocal
space, the gaussian hamiltonian becomes: 
\begin{eqnarray}
H_{B} &=&\frac{u}{2}\sum_{q\neq 0}K^{-1}\Pi _{q}\Pi _{-q}+Kq^{2}\Theta
_{q}\Theta _{-q}  \nonumber \\
&&+\frac{\pi u}{2L}\left( \frac{\widehat{Q}^{2}}{K}+K\widehat{J}^{2}\right) .
\label{qj}
\end{eqnarray}
We have split the hamiltonian into the phonon part and the non-bosonic zero
mode part. The first term can indeed be rewritten as: 
\begin{equation}
H_{phonon}=\sum_{q\neq 0}u\left| q\right| \left( b_{q}^{+}b_{q}+\frac{1}{2}%
\right)
\end{equation}
with the phonon operators: 
\begin{eqnarray}
b_{q} &=&\sqrt{\frac{K|q|}{2}}\left( \Theta _{q}-\frac{q}{K\left| q\right| }%
\Phi _{q}\right) ,  \label{p1} \\
b_{q}^{+} &=&\sqrt{\frac{K|q|}{2}}\left( \Theta _{-q}-\frac{q}{K\left|
q\right| }\Phi _{-q}\right) .  \label{p2}
\end{eqnarray}
The second term in the hamiltonian is standard in conformal field theory; it
corresponds to finite size corrections to the energy when one adds particles
or creates persistent currents in the Luttinger liquid. The corresponding
states are built by means of Haldane's $U_{\pm }$ operators which act as ladder
operators in Fock space\cite{haldane79}. This $(Q,J)$ part of the
hamiltonian is often called in CFT a zero mode part. The corresponding
excitations {\it may} however carry momentum. A non zero $J$ excitation
creates indeed a persistent current with momentum $Jk_{F}$. These states are
therefore non-dispersive since their momentum may only assume the discrete
values $Jk_{F}$.

The spectrum of the hamiltonian results from a convolution of plasmon
excitations and of these $(Q,J)$ excitations as is apparent in figures (\ref
{fig1},\ref{fig2}): two linear plasmon branches rise from each local
minimum of the energy obtained for the zero-mode states $(Q,J)$.
 It is important to note that there are selection rules on the
allowed values of $(Q,J)$, which refer back to the quantum statistics of the
particles: as reviewed in the previous section, the gaussian model can be
considered either for bosons or fermions, which results in different
bosonization formulas. For bosons, $J$ is constrained to be an even integer
while for fermions, $Q$ and $J$ must have the same parity. This then leads
to two different spectra as can be seen from figures (\ref{fig1}) and (\ref
{fig2}): for instance, for bosons the state $(Q=1,J=0)$ is available while
it is forbidden for fermions; conversely $(Q=1,J=1)$ is available to fermions
but not to bosons. Thus we have
two different theories: the same hamiltonian leads to different
properties depending on whether we consider
a Fock space of bosons or a Fock space of fermions\cite{hal81}. We will
call the LL with bosonic (resp. fermionic) selection rules: the bosonic
(resp. fermionic) LL. For the bosonic LL, as depicted in figure (\ref{fig1})
the spectrum in arbitrary charge sectors has the same form but for a shift
in energies: in the charge sector $Q$ one must add the constant $\pi
uQ^{2}/(2L)$ to the energy. The same energies are found for the fermionic LL
in charge sectors for which $Q$ is an even integer, but if $Q$ is an odd
integer there is a new spectrum with local minima at momenta $\pm k_{F}$
and not $k=0$ (figure \ref{fig2}).

In the rest of the paper we refer to this parametrization of the
spectrum in terms of phonons and zero modes as the zero mode basis; this is
to be distinguished from the quasiparticle basis which we will build later.
A property which will prove crucial for the rest of the discussion is the
fact that in the free-fermion case a quasiparticle basis exists as an
alternative to the zero mode basis: instead of the zero modes basis, it is
indeed possible to parametrize the spectrum in terms of the usual Landau
quasiparticles. Below we show that
a similar quasiparticle basis can be built in the interacting case. While
fractional quasiparticles do occur in exactly solvable models (the holon,
the spinon), scant contact had been made with the bosonization approach
as mentioned earlier. In the low-energy limit, using
the bosonization formalism, we will directly recover 
the fractional excitations predicted in
Bethe Ansatz, with the advantage that the simplifications brought by the
low-energy limit will allow a complete characterization, giving for
instance easy access
to statistical phases.

\section{Fractional excitations of the spinless Luttinger Liquid.}

\label{ch-two}

This section is divided as follows: first, we discuss the property of
chiral separation which is central to the physics of fractionalization;
then, we exhibit fractional quasiparticles for the bosonic LL
before turning to the fermionic LL for which we will find a different set of
elementary excitations.

\subsection{Chiral separation.}

\label{sec2-1}

\subsubsection{Chiral vertex operators and fractionalization.}

\label{cvo}

The gaussian model is endowed with a very basic property which is that of chiral
separation, i.e. we can split it into two commuting parts corresponding to
right or left propagation of the fields. This is a property which is
systematically used by CFT in the analysis of conformally invariant systems.
Indeed: 
\begin{equation}
H_{B}=\frac{u}{2}\int_{0}^{L}dx\ K^{-1}\Pi (x)^{2}+K(\nabla \Theta (x))^{2},
\end{equation}
\begin{equation}
\Rightarrow \left[ \partial _{x}^{2}-\frac{1}{u^{2}}\partial _{t}^{2}\right]
\Theta (x,t)=0
\end{equation}
More precisely we introduce the following chiral fields: 
\begin{equation}
\Theta _{\pm }(x)=\Theta (x)\mp \frac{\Phi (x)}{K}
\end{equation}
which are related to the phonon operators by: 
\begin{eqnarray}
q &>&0:\;b_{q}=\sqrt{\frac{K|q|}{2}}\Theta _{+,q}, \\
q &<&0:\;b_{q}=\sqrt{\frac{K|q|}{2}}\Theta _{-,q}.
\end{eqnarray}
In terms of these fields the hamiltonian becomes:

\begin{eqnarray}
H_{B} &=&H_{+}+H_{-}, \\
H_{\pm } &=&\frac{uK}{4}\int_{0}^{L}dx\;:\left( \partial _{x}\Theta _{\pm
}(x)\right) ^{2}: \\
&=&\sum_{\pm q>0}u\left| q\right| :b_{q}^{+}b_{q}:+\frac{\pi u}{LK}\left( 
\frac{\widehat{Q}\pm K\widehat{J}}{2}\right) ^{2}.
\end{eqnarray}
$H_{+}$ only contains right-moving phonons and similarly for $H_{-}$ with
left-moving phonons. It is clear also that $\left[ H_{+},H_{-}\right] =0$.
Let us show now that these fields $\Theta _{\pm }$ are chiral; they obey the
equal-time commutation relations: 
\begin{equation}
\left[ \Theta _{\pm }(x),\mp \frac{K}{2}\partial _{y}\Theta _{\pm
}(y)\right] =i\delta (x-y),
\end{equation}
which implies that the momentum canonically conjugate to $\Theta _{\pm }$ is 
$\Pi _{\Theta _{\pm }}=\mp \frac{K}{2}\partial _{x}\Theta _{\pm }$. The
equations of motions for these fields are: 
\begin{equation}
u\partial _{x}\Theta _{\pm }=\mp \partial _{t}\Theta _{\pm }.
\end{equation}
Thus: $\Theta _{\pm }(x,t)=\Theta _{\pm }(x\mp ut)$ which means that we have
chiral fields indeed. The superfluid phase has therefore been parametrized
as: $\Theta (x,t)=\frac{1}{2}\left[ \Theta _{+}(x-ut)+\Theta
_{-}(x+ut)\right] $.

One may define chiral density operators as well as the corresponding chiral
charges as: 
\begin{eqnarray}
\rho _{\pm }(x) &=&\frac{1}{2\sqrt{\pi }}\partial _{x}\Phi _{\pm }(x)=\frac{%
\delta \rho (x)\pm Kj(x)}{2},  \label{chiral} \\
\widehat{Q}_{\pm } &=&\frac{\widehat{Q}\pm K\widehat{J}}{2}.
\end{eqnarray}
Those chiral densities obey the anomalous (Kac-Moody) commutation relations: 
\begin{equation}
\left[ \rho _{\pm }(x),\rho _{\pm }(y)\right] =\mp \frac{iK}{2\pi }\partial
_{x}\delta (x-y).
\end{equation}

Let us now consider the injection of $Q$ particles with a momentum $q$
and current $J$. In that case, the plasmon total momentum is equal to
$q-J(k_{F}+\frac{\pi Q}{L})$. In the bosonization formalism, the operator creating
 this state is:
\begin{equation}
V_{Q,J}(q)=\frac{1}{\sqrt{L}}\int_{0}^{L}dxe^{i(q-Jk_{F})x}:\exp -i\sqrt{\pi 
}(Q\Theta -J\Phi ):.
\end{equation}
This can also be rewritten as: 
\begin{eqnarray}
V_{Q,J}(q) &=&\frac{1}{\sqrt{L}}\int_{0}^{L}dxe^{i(q-Jk_{F})x}\exp -i\sqrt{%
\pi }Q_{+}\Theta _{+}(x)  \nonumber \\
&&\times \exp -i\sqrt{\pi }Q_{-}\Theta _{-}(x).
\end{eqnarray}
As a function of time: 
\begin{eqnarray}
V_{Q,J}(q,t) &=&\frac{1}{\sqrt{L}}\int_{0}^{L}dxe^{i(q-Jk_{F})x}\exp -i\sqrt{%
\pi }Q_{+}\Theta _{+}(x-ut)  \nonumber \\
&&\times \exp -i\sqrt{\pi }Q_{-}\Theta _{-}(x+ut).
\end{eqnarray}
There is therefore a splitting into two counter-propagating states. For
non-interacting electrons the chiral charges $Q_{\pm }$ are integers since $%
K=1$ and the operators $\exp -i\sqrt{\pi }Q_{\pm }\Theta _{\pm }(x\mp ut)$
are just those of $Q_{\pm }$ Landau quasiparticless. But in the general case
this is not true anymore: we will therefore have states carrying fractional
charges.

We now define the chiral vertex operators which appeared in the previous
expression as: 
\begin{equation}
V_{Q_{\pm }}^{\pm }(x)=\exp -i\sqrt{\pi }Q_{\pm }\Theta _{\pm }(x),
\end{equation}
where the upperscript $\pm $ refers to the direction of propagation. They
obey the following commutation rules:

\begin{eqnarray}
\left[ \rho (x),V_{Q_{\pm }}^{\pm }(y)\right] &=&Q_{\pm }\delta
(x-y)\;V_{Q_{\pm }}^{\pm }(x), \\
\left[ \widehat{Q},V_{Q_{\pm }}^{\pm }(x)\right] &=&Q_{\pm }\;V_{Q_{\pm
}}^{\pm }(x), \\
\left[ \widehat{J},V_{Q_{\pm }}^{\pm }(x)\right] &=&\frac{Q_{\pm }}{K}%
\;V_{Q_{\pm }}^{\pm }(x),
\end{eqnarray}
which shows they carry charges $Q_{\pm }=\frac{Q\pm KJ}{2}$ which are
non-integral in general. The above operator identity means that the charge
is 'sharp': by 'sharp' we mean that the charge found is not a quantum
average ( $<Q>$ is not necessarily quantized of course). This is a point we
want to stress because this means that these quantum states are genuinely
fractional. This shows then that if one injects $Q$ particles with current $%
J $ in a LL, one should observe a charge $Q_{+}=\frac{Q+KJ}{2}$ state
propagating to the right at velocity $u$ and a charge $Q_{-}=\frac{Q-KJ}{2}$
going to the left with velocity $-u$. For instance, let us inject an
electron exactly at the right Fermi point: this is a $(Q=1,J=1)$ excitation
(with no plasmon excited); there would then be fractionalization into a
charge $\frac{1+K}{2}$ state going to the right and a charge $\frac{1-K}{2}$
going to the left.

The most important property of these fractional states is that they are
exact eigenstates of the gaussian hamiltonian. The proof requires a proper
definition of their Fourier transform because they are anyons, as will be
shown shortly: from equation (\ref{anyon}) it is clear indeed
that the commutation relations are anyonic with an anyonic phase 
\begin{equation}
\theta =\pm \pi \frac{Q_{\pm }^{2}}{K}.
\end{equation}
Due to its anyonic character $V_{Q_{\pm }}^{\pm }(x)$ does not obey periodic
boundary conditions; if we use the expressions of the fields $\Phi $ and $%
\Theta $ (equations \ref{phi}, \ref{theta}), we immediately find that:

\begin{equation}
V_{Q_{\pm }}^{\pm }(x+L)=\exp \pm i2\pi \frac{Q_{\pm }^{2}}{K}\;V_{Q_{\pm
}}^{\pm }(x).
\end{equation}
The Fourier transform is then defined as: 
\begin{equation}
V_{Q_{\pm }}^{\pm }(q_{n})=\frac{1}{\sqrt{L}}\int_{0}^{L}dx\exp -i\left( 
\frac{2\pi }{L}n\pm \frac{2\pi }{L}\frac{Q_{\pm }^{2}}{K}\right)
x\;V_{Q_{\pm }}^{\pm }(x),
\end{equation}
with a pseudo-momentum $q_{n}$ quantized as: 
\begin{eqnarray}
q_{n} &=&\frac{2\pi }{L}n\pm \frac{2\pi }{L}\frac{Q_{\pm }^{2}}{K} \\
&=&\overline{q_{n}}\pm \frac{2\pi }{L}\frac{Q_{\pm }^{2}}{K},
\end{eqnarray}
(where we have defined a phonon part $\overline{q_{n}}$ of the momentum).

{The operators $V_{Q_{\pm }}^{\pm }(q_{n})$  are such that:

1)  $V_{Q_{\pm }}^{\pm }(q)|\Psi _{0}>$ 
 is an exact eigenstate of the
chiral hamiltonian $H_{\pm }$ with energy: 
\begin{equation}
E(Q_{\pm },\overline{q_{n}})=\left[ u\left| \overline{q_{n}}\right| +\frac{%
\pi u}{2L}\frac{Q_{\pm }^{2}}{K}\right] .
\end{equation}
where $|\Psi _{0}>$ is the interacting ground state (see appendix).}
It has a linear dispersion.

2) The states {created by the $V_{Q_{\pm }}^{\pm }(q_{n})$
to which one adds the phonon
excitations form a complete set}. This is obvious because the states $%
V_{Q,J}(x)$ span the full Fock space\cite{basis}.

\subsubsection{The LL spectrum in terms of fractional quasiparticles.}

Let us consider figures (\ref{fig1},\ref{fig2}) which show the spectrum of
the LL hamiltonian in various charge sectors {and ask the
following question}: what happens
when one adds $Q$ particles to the system (i.e. in the charge sector $Q$)?
In the standard view of the LL spectrum based on the phonon and zero modes
basis, the dynamics of the charge added to the LL is unclear because it is
concealed in the zero modes. The parametrization of the spectrum in terms of
the zero modes and the phonons does not allow to find {what happens once the charge 
$Q$ is } added to the system because that choice of basis involves the use of
non-dynamical states (Haldane's $U_{pm}$ operators, which describe the zero
modes). {By} contrast the quasiparticle basis only involves states which have
a dynamics (the phonons and the fractional states) and we are therefore able
to tell what happens to the charge, how much of it will move to the right,
and so forth: if we consider the two branches starting from $k=Jk_{F}$ in
the charge sector $Q$, the right branch corresponds to a right-moving
fractional excitation with linear dispersion and with charge $Q_{+}=\frac{%
Q+KJ}{2}$, while the left branch is due to a left-moving fractional state
with charge $Q_{-}=\frac{Q-KJ}{2}$. The continuum in between the branches
{simply} results  from the creation of the two fractional excitations with
both non-zero momentum $\overline{q_{+,n}}$ and $\overline{q_{-,n}}$ (on the
right branch, a charge $Q_{-}=\frac{Q-KJ}{2}$ is also created but it has
zero momentum $\overline{q_{-,n}}=0$, and conversely on the left branch).

The direct way to find out how the charge $Q$ will behave, is to exhibit the
quantum states which will describe the propagation of the charge. This is
what the quasiparticle basis does because it directly considers the states
involved in the dynamics of the charge. Of course, the two bases (the
quasiparticle basis and the zero mode basis) are mathematically equivalent
and therefore lead to identical physics: therefore the charge dynamics can
also in principle be determined in the zero mode basis, but in the
quasiparticle basis, we have the benefit that the spectroscopy
immediately {tells us} the fate of the charge added to the system. In sharp
contrast, in the zero mode basis, the spectroscopy is not useful because the
states used in that basis are the phonons (which have no charge) and the $U_{\pm}$
operators (which have charge but no dynamics). We will give such an argument in the
next section: this will prove in an independent manner the fractionalization
of the LL spectrum (in a way which does not depend on the explicit
construction of the fractional states operators).

\subsubsection{Selection rules and fractionalization.}

\label{frac-current}

The fractional charges carried by the fractional excitations considered
above are not arbitrary: they must take { on} the values 
\begin{equation}
Q_{\pm }=\frac{Q\pm KJ}{2},  \label{spec-charge}
\end{equation}
where both $Q$ and $J$ are integers. We may view these constraints on the
allowed spectrum of fractional charges as selection rules. These selection
rules have however a clear physical meaning which we discuss now.

Although {these excitations do not carry the electron quantum numbers because of the
fractionalization of the spectrum, nevertheless the elementary constituents of our system are
electrons (they are the high energy elementary particles of our systems)}: this
means that they alone define the structure of Fock space, with the
implication that all physical states must consist {of} an integrer number of
electrons.
 {Despite the fact that there are} fractional states, the previous remark
implies that these fractional states will be created in appropriate
combinations so that the total charge is always an integer. This is the
explanation of the previous selection rules we found, which {\it in fine}
{enforce} the basic constraint that {we started out with}
electrons. We may view these selection rules as being topological since they
are directly related to the structure of the Fock space.

It is easy to show that eq. (\ref{spec-charge}) {immediately follows from the requirement} that all
states are electronic. We consider two counterpropagating states with
arbitrary charge $Q_{+}$ and $Q_{-}$; we make no hypothesis on the values of
the charges,
nor on the nature of the chiral states (we do {\it not} assume they
correspond to $V_{Q_{+}}^{+}$ and $V_{Q_{-}}^{-}$). The only assumptions we
make are the following: (a) the one-dimensionality which means that the
eigenstates have momenta in one of either two directions and (b) that
the current density operator is renormalized. We then have two constraints
on the values that the charges $Q_{+}$ and $Q_{-}$ may assume: since our
Fock space is that of electrons, {all the states contain an integer number} of
electrons i.e. $Q_{+}+Q_{-}=Q$ is an integer. The second constraint stems from the
renormalization of the current density operator: 
\begin{equation}
j_{R}(x)=uKj(x)=uK\left( \frac{1}{\sqrt{\pi }}\partial _{x}\Theta (x)\right)
\end{equation}
where $j(x)$ is the current density in the non-interacting case. This
expression can be derived from the continuity equation\cite{haldane79}.
Going around a ring of length $L$ in the LL we get a (persistent) current
which must be quantized: 
\begin{equation}
J_{R}=\int_{0}^{L}dxj_{R}(x)=uKJ
\end{equation}
where $J$ is an integer\cite{f4}; but the current carried by the states with
charges $Q_{+}$ and $Q_{-}$ is $J_{R}=u(Q_{+}-Q_{-}).$ Therefore: 
\begin{equation}
(Q_{+}-Q_{-})=KJ,\;J\;%
\mathop{\rm integer}%
\end{equation}
while: 
\begin{equation}
(Q_{+}+Q_{-})=Q,\;Q\;%
\mathop{\rm integer}%
\end{equation}
Solving for these constraints, one recovers the selection rules {(\ref{spec-charge}), i.e the spectrum of
fractional charges}. We observe in passing that this argument {\it does not}
depend on our formal algebraic derivation of subsection (\ref{cvo}) and
{provides an alternative proof of} the existence of fractional states as well as it yields
the allowed charge spectrum. In that argument, {\it fractionalization
follows from the renormalization of the current in the presence of
interactions}.

\subsection{Elementary excitations of the bosonic LL.}

\label{sec2-2}

\subsubsection{Elementary excitations.}

We now establish a series of new results concerning the elementary chiral
excitations of a non-chiral LL. We would
like to find a basis of elementary excitations, i.e. identify objects from
which all the other excitations can be built. It will be useful to use a
spinor notation to represent the fractional states: 
\begin{equation}\label{toto}
\left( 
\begin{tabular}{l}
$Q_{+}$ \\ 
$Q_{-}$%
\end{tabular}
\right) =\left( 
\begin{tabular}{l}
$\frac{Q+KJ}{2}$ \\ 
$\frac{Q-KJ}{2}$%
\end{tabular}
\right) .
\end{equation}
(\ref{toto}) should be understood as follows: the fractional state $%
V_{Q_{+}}^{+}$ which is an anyon propagating with velocity $u$ is created
along with the fractional state $V_{Q_{-}}^{-}$ which propagates in the
{opposite} direction with the velocity $-u$. The selection rules are
encoded in the second spinor: the equation is then read as meaning that
addition of $Q$ particles with (persistent) current $J$ will result in a
splitting into the two counterpropagating fractional states $V_{Q_{+}}^{+}$
and $V_{Q_{-}}^{-}$.

We must carefully distinguish between Bose and Fermi statistics because of
the constraints on $Q$ and $J$. Let us consider bosons first: since $J$ is
even we can rewrite it as $J=2n$ where $n$ is now an arbitrary integer. But
then for {\it bosons this implies that the spinor can be written in terms of
two other independent spinors}:

\begin{eqnarray}
\left( 
\begin{tabular}{l}
$Q_{+}$ \\ 
$Q_{-}$%
\end{tabular}
\right) &=&\left( 
\begin{tabular}{l}
$\frac{Q+KJ}{2}$ \\ 
$\frac{Q-KJ}{2}$%
\end{tabular}
\right)  \nonumber \\
&=&Q\left( 
\begin{tabular}{l}
$\frac{1}{2}$ \\ 
$\frac{1}{2}$%
\end{tabular}
\right) +n\left( 
\begin{tabular}{l}
$K$ \\ 
$-K$%
\end{tabular}
\right) .  \label{bose-q}
\end{eqnarray}

This implies that in real space the fractional charge excitation is: 
\begin{equation}
V_{Q_{\pm }}^{\pm }(x)=\left[ V_{1/2}^{\pm }(x)\right] ^{Q}\left[ V_{\pm
K}^{\pm }(x)\right] ^{n},
\end{equation}
where $Q$ and $n$ are now {\it independent integers} of arbitrary sign: $%
(Q,n)\in Z^{2}$. In reciprocal space, one has a convolution for the exact
fractional eigenstate: 
\begin{eqnarray}
V_{Q_{\pm }}^{\pm }(\overline{q}) &=&\int ..\int \prod_{i=1}^{Q}d\overline{%
q_{i}}\left[ V_{1/2}^{\pm }(\overline{q_{i}})\right] \prod_{j=1}^{n}d%
\overline{p_{j}}\left[ V_{\pm K}^{\pm }(\overline{p_{j}})\right]  \nonumber
\\
&&\times \delta (\sum_{i=1}^{Q}\overline{q_{i}}+\sum_{j=1}^{n}\overline{p_{j}%
}-\overline{q})
\end{eqnarray}
(the momenta in that expression are the phonon parts $\overline{q_{n}}$ of
the momentum of the operator: for $V_{Q_{\pm }}^{\pm }(q_{n}),$ $q_{n}=%
\overline{q_{n}}\pm \frac{2\pi }{L}\frac{Q_{\pm }^{2}}{K}$ and $\overline{%
q_{n}}=\frac{2\pi n}{L}$)\cite{f6}.

The above equation demonstrates clearly that the excitation $V_{Q_{\pm
}}^{\pm }(\overline{q})$ can be built from $Q$ charge $1/2$ states $%
V_{1/2}^{\pm }$ and $n$ charge $\pm K$ states $V_{\pm K}^{\pm }$. The whole
spectrum of fractional excitations is thus built by repeated creation of $%
V_{1/2}^{\pm }$ and $V_{\pm K}^{\pm }$ which means that they are the {\it %
elementary excitations }we were seeking. These two elementary excitations
will be identified in the following as {respectively} the spinon (for spin systems) and a
(1D) Laughlin quasiparticle.

\subsubsection{Wavefunctions of the fractional excitations.}

\label{sec-222}

{To be complete, we} compute the wavefunctions of all the chiral
excitations. We will first need the ground state wavefunction which is
simply a Jastrow wavefunction: this is of course expected since the gaussian
hamiltonian is the 1D version of the acoustic hamiltonian of Chester and
Reatto's Jastrow theory of He4\cite{chester} and is also
identical to Bohm-Pines RPA plasmon hamiltonian adapted to 1D\cite
{bohm-pines}. Since the gaussian hamiltonian is a sum of
oscillators, the ground state is a gaussian function of the densities:

\begin{eqnarray}
&&\Psi _{0,B}(\{\rho _{q}\})  \nonumber \\
&=&\exp (-\sum_{q\neq 0}\frac{\pi }{2K\left| q\right| }\rho _{q}\rho _{-q})
\\
&=&\exp \frac{1}{2K}\int \int dxdx^{\prime }\;\widehat{\rho }(x)\ln \left|
\sin \frac{\pi }{L}(x-x^{\prime })\right| \widehat{\rho }(x^{\prime }).
\end{eqnarray}
This expression is valid for the bosonic LL; for the fermionic LL,
antisymmetry is recovered by observing that the fermionic LL {simply}
derives from the bosonic LL through a singular gauge transformation on the
bosonic LL (the Jordan-Wigner transformation)\cite{f8}: this is exactly as
in the composite boson Chern-Simons theory for which the hamiltonian is
plasmon-like at the one-loop level (RPA) and whose ground state is of course
symmetric (the modulus of Laughlin wavefunction); in that theory, the
Laughlin state is then found after undoing the Chern-Simons gauge
transformation\cite{kiv}. Similarly undoing the Jordan-Wigner transformation
amounts to multiplying the bosonic ground state by the phase factor $%
\prod_{i<j}sgn(x_{i}-x_{j})=\prod_{i<j}\left( \frac{(x_{i}-x_{j})}{\left|
x_{i}-x_{j}\right| }\right) $ (that phase factor is found by applying the
Jordan-Wigner operator on the ground state). The wavefunction is {the 1D analog
of} the 2D Laughlin state of FQHE if we rewrite the previous expression in
terms of the particles' positions: $\rho (x)=\sum_{i}\delta (x-x_{i})$, and
by introducing the circular coordinates $z=\exp i\frac{2\pi }{L}x$: 
\begin{equation}
{\psi }_{0,B}(\{x_{i}\})=\prod_{i<j}\mid z_{i}-z_{j}{\mid }^{1/K}.
\label{bosonic}
\end{equation}

The {wavefunctions of the excited states} can now be computed\cite{f9}. Let us consider
first the operator $\exp -i\sqrt{\pi }\alpha \Theta (x_{0})$; since $\Theta $
is the canonical conjugate of the field $\Pi =\partial _{x}\Phi =-\sqrt{\pi }%
\delta \widehat{\rho }$, 
\begin{equation}
\Theta (x)=-\frac{1}{i\sqrt{\pi }}\frac{\delta }{\delta \widehat{\rho }(x)},
\end{equation}
and therefore: 
\[
\exp -i\sqrt{\pi }\alpha \Theta (x_{0})|\Psi _{0,B}> 
\]
\begin{eqnarray}
&=&\exp \alpha \frac{\delta }{\delta \widehat{\rho }(x_{0})}  \nonumber \\
&&\times \exp \frac{1}{2K}\int \int dxdx^{\prime }\;\widehat{\rho }(x)\ln
\left| \sin \frac{\pi }{L}(x-x^{\prime })\right| \widehat{\rho }(x^{\prime })
\nonumber \\
&=&\exp \frac{\alpha }{K}\int dx\;\widehat{\rho }(x)\ln \left| \sin \frac{%
\pi }{L}(x-x_{0})\right| \;\Psi _{0,B}  \nonumber \\
&=&C\prod_{i}\left| z_{i}-z_{0}\right| ^{\alpha /K}\prod_{i<j}\left|
z_{i}-z_{j}\right| ^{1/K}
\end{eqnarray}
where $C$ is an unessential constant.

Similarly:

\begin{eqnarray}
&&\exp \pm i\sqrt{\pi }\frac{Q_{\pm }}{K}\Phi (x)\;\exp \mp i\frac{Q_{\pm }}{%
K}k_{F}x \nonumber \\ 
&=&\exp\mp i\pi \frac{Q_{\pm }}{K}\int_{0}^{x}\widehat{\rho }(y)dy \nonumber\\
&=&\exp \mp i\pi \frac{Q_{\pm }}{K}\int_{0}^{L}\widehat{\rho }(y)\theta
(x-y)dy \nonumber \\
&=&\prod_{i}\left[ \frac{(x_{i}-x)}{\left| x_{i}-x\right| }\right] ^{\mp
Q_{\pm }/K} \nonumber \\
\end{eqnarray}
\begin{equation}
=\prod_{i}\left[ \frac{(z_{i}-z)}{\left| z_{i}-z\right| }\right] ^{\mp
Q_{\pm }/K}\exp \pm ik_{F}\frac{Q_{\pm }}{K}\left( \frac{\sum_{i}x_{i}}{N_{0}%
}+x\right) ,
\end{equation}
where we {use the definitions of $\rho$ and $z$ introduced 
above and where} 
\begin{equation}
\theta (x)=\frac{1}{i\pi }\ln \left[ \frac{-x}{\left| x\right| }\right] ,
\end{equation}
is the Heaviside step function.

 The above operator can thus be 
seen as a generalized Jordan-Wigner operator, since it multiplies
wavefunctions by a singular phase factor; in this manner we recover the phase
$\prod_{i<j}\left( \frac{(x_{i}-x_{j})}{\left|
x_{i}-x_{j}\right| }\right) $ of the ground state of the fermionic LL. Finally we have that: 
\begin{eqnarray}
&&V_{Q_{+}}^{+}(x)\Psi _{0,B}(x_{1},..,x_{N})  \nonumber \\
&=&C\prod_{i}(\overline{z_{i}}-\overline{z})^{Q_{+}/K}\;\prod_{i<j}\left|
z_{i}-z_{j}\right| ^{1/K}  \nonumber \\
&&\times \exp ik_{F}\frac{Q_{+}}{2K}\left( \frac{\sum_{i}x_{i}}{N_{0}}%
+x\right) ,
\end{eqnarray}
with a similar expression for $V_{Q_{-}}^{-}$ (the bar over $z$ denotes
complex conjugation). It is noteworthy that these wavefunctions are
obtained by multiplying a Jastrow ground state with a Laughlin-like
prefactor $\prod_{i}\left| z_{i}-z\right| ^{Q_{\pm }/K}$ which generalizes
the Laughlin quasihole factor $\prod_{i}(z_{i}-z)$. We can now write down
the wavefunctions of the two elementary excitations: 
\begin{eqnarray}
&&V_{1/2}^{+}(x)\Psi _{0,B}(x_{1},..,x_{N})  \nonumber \\
&=&C\prod_{i}(z_{i}-z)^{1/2K}\;\prod_{i<j}\left| z_{i}-z_{j}\right| ^{1/K} 
\nonumber \\
&&\times \exp -i\frac{k_{F}}{2K}\left( \frac{\sum_{i}x_{i}}{N_{0}}+x\right) ,
\end{eqnarray}
and,
\begin{eqnarray}
&&V_{K}^{+}(x)\Psi _{0,B}(x_{1},..,x_{N})  \nonumber \\
&=&C\prod_{i}(z_{i}-z)\;\prod_{i<j}\left| z_{i}-z_{j}\right| ^{1/K} 
\nonumber \\
&&\times \exp -ik_{F}\left( \frac{\sum_{i}x_{i}}{N_{0}}+x\right) .
\end{eqnarray}
We see that $V_{K}^{+}(x)$  is nothing but the 1D counterpart of the 2D Laughlin quasi-hole wavefunction, provided we
make the following correspondence between 1D and 2D wavefunctions: $%
K\Longleftrightarrow \nu $, $z=\exp i2\pi x/L\Longleftrightarrow z=x+iy$ (up
to a galilean boost absorbing the factor $\exp -ik_{F}\left( \frac{%
\sum_{i}x_{i}}{N_{0}}+x\right) $): {\it in view of the formal analogy we
will call that state a 1D Laughlin state}.

\subsubsection{The spinon.}

\label{sec-223}

We found an elementary excitation $V_{1/2}^{\pm }(x)$ for the bosonic LL
carrying a charge $1/2$. When we consider spins, which are
hard-core bosons, this result translates into having a state carrying a spin 
$\Delta S_{z}=1/2$ with respect to the ground state. In spin language,
adding a particle into the system $\left( Q=1\right) $ corresponds to
flipping a spin $\left( \Delta S_{z}=1\right) $. But it follows from eq.(\ref
{bose-q}) that this excitation is a composite of two
elementary excitations, each carrying a charge $1/2$. Therefore a pair of
states with spin $S_{z}=1/2$ is created when one flips a spin $\left( \Delta
S_{z}=1\right) $. We naturally identify this fractional spin excitation as a
spinon. The spinon can be generated without any Laughlin quasiparticless if
the spin current is zero $\left( J=0\right) $: this is a process which we
{term} a {\it pure spin process}, to be distinguished from a {\it %
pure spin current process} $(S_{z}=0)$ which 
generates Laughlin quasiparticle-quasihole pairs (see below).

The properties of the spinon specifically depend on the LL parameter $K$.
Although the spin is always $S_{z}=1/2$ the exchange statistics varies
continuously with $K$ (i.e. when one varies the interaction): 
\begin{equation}
\theta _{spinon}=\frac{\pi }{4K}.
\end{equation}
For instance for $K=1/2$ (which corresponds to $SU(2)$ symmetric spin
interactions) the spinon is a semion. In that special case, the spinon
wavefunction we obtain coincides exactly with that proposed by Haldane for
the Haldane-Shastry spin chain\cite{hal94}: 
\begin{eqnarray}
\Psi _{spinon}(z) &=&\prod_{i}(z_{i}-z)\;\prod_{i<j}\left|
z_{i}-z_{j}\right| ^{2}  \nonumber \\
&&\times \exp -ik_{F}\left( \frac{\sum_{i}x_{i}}{N_{0}}+x\right)
\end{eqnarray}
In this expression the coordinates are those of the down spins. For 
$K=1/2$ the spinon and the spin $K$ Laughlin quasiparticles are
identical. {Although we have discussed fractional
excitations for spin systems}, the previous considerations apply of course to
bosons: the ''spinon'' is then a {\it charge} $1/2$ excitation. For
convenience we will call the excitation a spinon even when we consider
bosons.

\subsubsection{The LL Laughlin quasiparticle.}

The second elementary excitation we found has the following wavefunction: 
\begin{eqnarray}
\Psi _{Laughlin-qp}(z_{0}) &=&\prod_{i}(z_{i}-z_{0})\;\prod_{i<j}\left|
z_{i}-z_{j}\right| ^{1/K}  \nonumber \\
&&\times \exp -ik_{F}\left( \frac{\sum_{i}x_{i}}{N_{0}}+x_{0}\right) ,
\end{eqnarray}
which leads us to identify it with a Laughlin quasiparticles. The parallels
which can be drawn between the 2D Laughlin quasi-hole and the
Luttinger liquid Laughlin quasiparticles are indeed very strong. For
instance as in 2D one can use the plasma analogy to find the fractional
charge: 
\begin{eqnarray}
&&\left| \Psi _{Laughlin-qp}(z_{0})\right| ^{2} =\left|
\prod_{i}(z_{i}-z_{0})\prod_{i<j}\left| z_{i}-z_{j}\right| ^{1/K}\right| ^{2}
\nonumber \\
&=&\exp \frac{1}{K}\int \int dxdx^{\prime }\left[ \widehat{\rho }(x)+K\delta
(x-x_{0})\right]  \nonumber \\
&&\ln \left| \sin \frac{\pi }{L}(x-x^{\prime })\right| \left[ \widehat{\rho }%
(x^{\prime })+K\delta (x^{\prime }-x_{0})\right] .
\end{eqnarray}
The above expression clearly shows that the charge carried by the excitation is $K$ in
agreement with the direct algebraic determination (using the operator $%
V_{K}$). There are however several differences between the LL Laughlin
quasiparticles and its 2D famous counter-part; first, there is no
analyticity {requirement in the 1D problem, 
since we do not have to project into }the lowest Landau level: we
have two chiralities and the LL Laughlin quasi-electron is simply 
\begin{eqnarray}
\Psi_{Laughlin-qe}(z_{0}) &=&\prod_{i}(z_{i}-z_{0})^{-1}\;\prod_{i<j}\left|
z_{i}-z_{j}\right| ^{1/K}  \nonumber \\
&&\times \exp -ik_{F}\left( \frac{\sum_{i}x_{i}}{N_{0}}+x_{0}\right) .
\end{eqnarray}

Second, while topological quantization forces the 2D FQHE Laughlin
quasiparticles to have a {\it rational} charge, the {charge of the 1D LL Laughlin
quasiparticles can take on} {\it arbitrary} real positive
values, in particular {\it irrational}. This is a very startling property:
irrational spin had already been proposed for solitons in coexisting CDW-SDW
systems by B. Horowitz\cite{horowitz}, but in a sense this is perhaps less
surprising since in one dimension there is no quantization axis for spin
which can therefore take a continuum of values. We show below that the
Laughlin quasiparticles also exist for the fermionic LL; furthermore we
will find that for the fermionic LL there is another elementary excitation
which may have an irrational charge.

How are Laughlin quasiparticles created in a LL? They are generated
whenever $J\neq 0$; they are always created as quasiparticle-quasihole
pairs. In particular in pure current processes $(Q=0)$ no ''spinon'' is
created and we have only Laughlin quasiparticle-quasihole (qp-qh) pairs. For
a persistent current $J$ excitation with $Q=0$ it follows from the
expression $Q_{\pm }=\frac{Q\pm KJ}{2}$ that $J/2$ quasiparticle-quasihole
pairs are generated.

From the above analysis we now can give a physical interpretation to the renormalization of the
current density operator in the presence of interactions:

\begin{eqnarray}
j(x) &=&\frac{1}{\sqrt{\pi }}\partial _{x}\Theta (x), \\
&\longrightarrow &j_{R}(x)=\frac{uK}{\sqrt{\pi }}\partial _{x}\Theta (x).
\end{eqnarray}
The velocity $u$ has been normalized to the Fermi velocity so that $u=1$ in
the absence of interactions for fermions ($K=1$). 
We have found  that current excitations were due to
Laughlin quasiparticles. The natural explanation of the renormalization is
therefore that the current is no longer carried by Landau quasiparticles
but by Laughlin quasiparticles with velocity $u$ and charge $K$.

\subsubsection{The bosonic LL spectrum in terms of fractional elementary
excitations.}

For the bosonic LL we can now add the following precisions to the
description of the spectrum.

For $Q=0$ excitations (see figure (\ref{fig1})), {\it the continuum is due
to multiple Laughlin quasiparticle-quasihole pairs}: the right branch
starting at $k=2k_{F}$ corresponds to the propagation of a 1D Laughlin
quasielectron while the left branch is due to a Laughlin quasihole; in
between the two lines, we have a continuum generated by these two
excitations. More generally at $k=2nk_{F}$ where $n$ is an arbitrary
integer, the two branches create a continuum of $n$ Laughlin
quasiparticle-quasihole pairs. (Note that for $n=0$, which is an exceptional
case, we have multiphonon processes.) Therefore the spectrum in the zero
charge sector is {\it not} a Landau quasiparticle-quasihole pair continuum
except at the special value $K=1$ which describes indeed in the low
energy limit a gas of hard-core bosons.

In the charge sector $Q=1$, pairs of charge one-half excitations are
created: they correspond to the ''spinons'' of spin chains; the pairs are
superimposed on the previous Laughlin quasiparticle continuum: for instance
a $2k_{F}$ excitation generates a Laughlin quasiparticle-quasihole pair in
addition to the ''spinon'' pair.

{The Laughlin quasiparticle and the spinon are dual states for the bosonic
LL; by duality we mean electromagnetic duality which exchanges charge and
current processes. Indeed the ''spinon'' is associated with charge processes
while the Laughlin quasiparticles is due to current excitations. The duality
operation which maps a bosonic LL onto another bosonic LL is:$%
K=1/(4K^{\prime })$ with $\Theta =2\Phi ^{\prime }$ and $\Phi =\Theta
^{\prime }/2$; zero modes then transform as $J=2Q^{\prime }$ et $Q=J^{\prime
}/2$. With these relations, the selection rule remains bosonic ($J^{\prime }$
even) while the hamiltonian $H_{B}\left[ K,\Theta ,\Phi \right] =H_{B}\left[
K^{\prime },\Theta ^{\prime },\Phi ^{\prime }\right] $ retains a gaussian
form. It is clear then that $K=1/2$ is a self-dual point while $V_{K}$ and $%
V_{1/2}$ create dual quasiparticles. This is not true for the fermionic LL.}

\subsubsection{The $XXZ$ spin chain.}

Let us illustrate these results on the specific example of the anisotropic
Heisenberg $XXZ$ spin chain. The hamiltonian of the $XXZ$ spin chain with anisotropy $\Delta $,
after a bipartite rotation is: 
\begin{equation}
H\left[ J,\Delta \right] =J\sum_{i}\left\{ -\frac{1}{2}\left(
S_{i}^{+}S_{i+1}^{-}+S_{i}^{-}S_{i+1}^{+}\right) +\Delta
S_{i}^{z}S_{i+1}^{z}\right\} .
\end{equation}
{As $\Delta$ is varied}, one finds three phases: i) for $\Delta >1$ one
gets an Ising antiferromagnet {the twofold degenerate
groundstate of which leads to solitonic excitations
with spin one-half $1/2$ domain walls}; ii) for $\Delta <1$ one has an
Ising ferromagnet; iii) for -$1\leq \Delta \leq 1$ we have the so-called $XY$
phase: this is the Luttinger liquid phase we are interested in. The
isotropic Heisenberg chain with $SU(2)$ invariance corresponds to $\Delta =1$
. The Luttinger liquid parameter was determined exactly by Luther and
Peschel on the basis of a comparison with the Baxter model \cite{luther}: 
\begin{equation}
K(\Delta )=\frac{\pi }{2\arccos \left( -\Delta \right) }.
\end{equation}
The spectrum in the sector $\Delta S_{z}=1$ is shown in figure (\ref{fig3})
for the Heisenberg model; its linearization through bosonization is also
shown in figure.

{Given that a spin one-half can be mapped onto a hard-core boson:\cite{f12}
through the Holstein-Primakov transformation, we can transpose the results we found 
for the bosonic LL to the $XXZ$ spin chain.} 

{If we want to compare
the bosonization linearized spectrum to the exact one there are however two 
provisos:}
(a) we have to
shift the bosonization spectrum by a momentum $\pi $ : this is due to the
bipartite transformation one makes in the bosonization of the $XXZ$ spin
chain (in order to change the sign of the $XY$ term) and (b), there is a
Brillouin zone: therefore we have to identify momenta modulo $2\pi $ and
since the Fermi vector is $k_{F}=\pi /2,$ excitations with $J/2$ odd (resp.
even) correspond to the same harmonics $k=\pi +Jk_{F}\equiv 0$ (resp. $\pi $%
). {Taking (a) and (b) into account}
, we can use the results of the previous section pertaining to the bosonic LL, to
recover the linearized spectrum of the $XXZ$ chain. 

We first consider the spin sector $\Delta S_{z}=0$. In figure (\ref{fig3})
starting from momentum $\pi $ we have two straight lines corresponding to
left and right moving phonons, {bounding} a continuum; due to the folding of
the continuum spectrum of the bosonic LL, one superimposes on these lines
the lines due to the creation of any even number of Laughlin qp-qh pairs
(the qp dispersion being given by one line, and that of the qh by the other;
if the qp is right-handed, its dispersion is that of the right line,
etc...). Similarly the lines starting from momentum zero or $2\pi $
correspond to the creation of an odd number of Laughlin qp-qh pairs. The
continuum is therefore seen to be parametrized entirely in terms of the
phonons and Laughlin quasiparticle-quasihole pairs while the zero mode basis
relies on phonons and zero modes. The $\Delta S_{z}=1$ continuum is
described in a similar manner but for the substitution of the phonons by a
pair of counterpropagating spinons. In the special case of $SU(2)$ symmetry,
the Laughlin quasiparticle and the spinon become identical operators. The
previous parametrization reduces then to one involving only pairs of spinons
because a pair of counterpropagating spinon plus a pair of
counterpropagating spinon-antispinon is equivalent to a pair of spinons
propagating in arbitrary directions. One then recovers the Bethe Ansatz
result.

{\it In the low-energy limit, we can now answer the various questions raised
in the introduction about the spectrum of the }$XXZ${\it \ chain:}

{\it -what is the nature of the }$\Delta S_{z}=1${\it \ continuum? It is
indeed a spinon pair continuum; but superimposed on them Laughlin
quasiparticle-quasihole pairs can exist. The spinon changes when the
anisotropy is varied: it acquires a statistical phase }$\pi /4K=\arccos
(-\Delta )/2${\it . Therefore the spinons at different anisotropy are not
adiabatically connected: they are orthogonal states;}\cite{f13}{\it \ this
is consistent with numerical computations of the spectral density of the }$%
SU(2)${\it \ spinon, where it is found that the }$SU(2)${\it \ spinon has a
zero quasiparticles weight for the }$XY${\it \ chain}\cite{talstra}{\it .}

{\it -what is the nature of the }$\Delta S_{z}=0${\it \ continuum? It is a
Laughlin qp-qh pair continuum with an unquantized spin }$S_{z}=\pm K=\pm \pi
/(2\arccos \left( -\Delta \right) )${\it ; in the }$SU(2)${\it \ symmetric
case, they are identical to the spinons. In the }$XY${\it \ limit, one
recovers the standard spin one continuum predicted through a Jordan-Wigner
transformation (}$K=1,${\it \ }$S_{z}=\pm 1${\it ). But in between these two
points, the elementary excitation is neither a spinon nor a Jordan-Wigner
fermion.}

\subsection{The fermionic Luttinger Liquid.}

\label{sec2-3}

\subsubsection{Elementary excitations : the Laughlin quasiparticles and the
''hybrid state''.}

\label{sec-231}

We now turn to fermions; the analysis of the elementary excitations will
differ from that found for the bosonic LL because the allowed ($Q,J$) states
obey different selection rules, namely $J$ is not constrained any more to be
an even integer, but must have the same parity as $Q$. We may therefore
write $Q-J=2n$. Then for {\it fermions} using eq.\ref{spec-charge}: 
\begin{equation}
\left( Q_{+},Q_{-}\right) =Q\;\left( \frac{1+K}{2},\frac{1-K}{2}\right)
-n\;\left( K,-K\right) .  \label{first}
\end{equation}

{The most general excitation once again, is built by applying 
$Q$ times $V_{\frac{1\pm K}{2}}^{\pm }$ and/or $n$ times $V_{\pm K}^{\pm
}$ to the ground state }; this means that we have identified a set of elementary
excitations for the fermionic LL. {Here too} we find Laughlin quasiparticles 
$V_{\pm K}^{\pm }$, but {instead of} the spinon {we get a} ''hybrid state'':
this is a consequence of statistics; as we will show below, that hybrid
state is self-dual and is intermediate between the Laughlin quasiparticle
and its dual state.

The Laughlin quasiparticle is created by current excitations: for
a pure current process ($Q=0,J\neq 0$) one indeed generates Laughlin qp-qh
pairs as the above equation shows. The continuum for zero charge excitations
($Q=0$) is often depicted as a (Landau) particle-hole continuum as in the
non-interacting system ($K=1$): this is incorrect; we have instead a
Laughlin quasiparticle-quasihole continuum. The latter does reduce to the
standard Landau quasiparticle continuum when $K=1$. For $k=Jk_{F}$ there is
a local minimum of the energy from which  two linear branches rise
corresponding to $J/2=-n$ pairs of Laughlin quasiparticles and quasiholes.
For the fermionic LL, the Laughlin quasiparticle is not the only state which
may have an irrational charge: this is also possible for the hybrid state.

{The hybrid state} is created in mixed charge and current
processes: this is the main difference with the bosonic LL for which the
decoupling between charge and current processes is complete. As reviewed in the introduction,
 there are
even-odd effects in the fermionic spectrum: the spectra {obtained by adding} an
even or an odd number of particles are qualitatively different for the
fermionic LL. The $Q=1$ continuum is understood as follows: the two branches
at $k_{F}$ correspond to a pair of hybrid excitations carrying a charge $%
\frac{1-K}{2}$ and $\frac{1+K}{2},$ {and propagating with velocities
respectively $-u$ and $u$} . 
At $-k_{F}$ the correspondence is reversed. More generally at $%
Jk_{F}$ ($J$ is an odd integer if $Q=1$), in addition to the hybrid
quasiparticles, one also creates $J/2-Q$ pairs of Laughlin quasiparticle and
quasihole. When $K=1$ the hybrid states reduce to  Landau quasiparticles.
It is interesting to note an evidence for these states in the work of
Safi and Schulz who considered the evolution of a charge $1$ wavepacket
injected at $k_{F}$ in a LL: they found that there was a splitting with an
average charge $<Q>=(1+K)/2$ propagating to the right and an average charge $%
<Q>=(1-K)/2$ going {to the left}\cite{safi}. This is exactly
{what we predict}. Note however a crucial difference: the charge
they find is a quantum average while we deal with elementary excitations
(exact eigenstates); this has an important consequence: while it is clear that
on average 
a charge may assume irrational values, our result goes
beyond that observation since it proves that there may exist in condensed
matter systems a genuine good quantum state with sharp irrational charge.

In this section, we have found that for the fermionic LL
there are two elementary excitations. One is the Laughlin quasiparticle
already found for the bosonic LL. The second one is a hybrid state
intermediate between the spinon and the Laughlin quasiparticle. The
excitations corresponding to $Q=0$ transitions (they are
particle-hole excitations in the non-interacting case), form
a Laughlin quasiparticle-quasihole pair continuum when $K\neq 1$.

\subsubsection{Dual basis and the dual quasiparticles.}

The elementary excitations we have derived form a basis from which all the
LL spectrum is recovered; by no means {is this choice of basis unique}:
other bases of elementary excitations are {generated with
matrices associated with a  basis change having} integer entries whose 
inverses are also
integer-valued: this ensures that all excitations are integral linear
combinations of the elementary excitations. (The matrices belong to $SL(2,Z)$%
.) For instance for fermions, another basis of elementary excitations consists 
of states $V_{\frac{1\pm K}{2}}^{\pm }$and $V_{1}^{\pm }$: 
\begin{equation}
\left( Q_{+},Q_{-}\right) =J\;\left( \frac{1+K}{2},\frac{1-K}{2}\right)
+n\;\left( 1,1\right) .  \label{dual}
\end{equation}
It is actually a dual basis  to the previous one: for fermions the
electro-magnetic duality which exchanges charge and current excitations
{ is expressed by} $K\longleftrightarrow 1/K$ and $\Phi
\longleftrightarrow \Theta $. This is a canonical transformation; it results
in: $H_{B}\left[ K,\Theta ,\Phi \right] =H_{B}\left[ 1/K,\Phi ,\Theta
\right] $. The fermionic selection rule $Q-J$ even is obviously preserved:
the transformation is therefore a duality operation for the fermionic LL.
Observe that the transformations differ from those for the bosonic LL.

What is the nature of the elementary excitation $V_{1}^{\pm }$? Under the
duality transformation $V_{K}^{\pm }\longrightarrow V_{-1}^{\pm }$ and $%
V_{-K}^{\pm }\longrightarrow V_{1}^{\pm }$. Therefore $V_{1}^{\pm }$ is an
excitation dual to the Laughlin quasiparticle. It carries a charge unity and
its wavefunction is: 
\begin{equation}
V_{1}^{+}(z_{0})\Psi _{F}=\prod_{i}(z_{i}-z_{0})^{1/K}\;\prod_{i<j}\left|
z_{i}-z_{j}\right| ^{1/K}\;\prod_{i<j}\frac{\left( z_{i}-z_{j}\right) }{%
\left| z_{i}-z_{j}\right| }
\end{equation}
We stress that although $V_{1}^{\pm }$ carries a unit charge, it is {\it not}
an electron: the statistical exchange phase is $\pi /K$ which means $%
V_{1}^{\pm }$ is an anyon ({it can be a fermion, in the special case}
 $K=1/(2n+1)$). The difference with the electron is quite clear
since the electron {creation operator} is:

\[
\Psi (x)=\sum_{n}\exp i(2n+1)k_{F}x 
\]
\begin{equation}
\exp i\left( \sqrt{\pi }\Theta (x)-\sqrt{\pi }(2n+1)\Phi (x)\right) ,
\end{equation}
while: 
\begin{equation}
V_{1}^{+}(x)=:\exp i\sqrt{\pi }(\Theta (x)-\frac{\Phi (x)}{K}):.
\end{equation}
{For $K\neq 1/(2n+1)$ the dual excitation appears to be a} non-linear soliton of the electron.
This excitation is interesting in many respects. If $K=1/(2n+1)$ (the
Laughlin fractions) the excitation is fermionic and the exchange statistics
of the operator is $\pi (2n+1)$. The dual quasiparticle corresponds then to
a sub-dominant harmonic of the electron Fourier expansion around $k\simeq
(2n+1)k_{F}$. If one attaches $2n$ flux tubes to the electron (i.e.
multiplies the electron operator by the Jordan-Wigner phase $\exp i\sqrt{\pi 
}2n\Phi $) the dual state becomes the dominant $k=k_{F}$ harmonics: this is
exactly the composite fermion construction and it may then be more fitting
to speak of a composite fermion (indeed, the statistics of the operator is  $%
(2n+1)\pi $ and not $\pi $). {Because of the similar
long distance behavior of their Green functions, Stone proposed to identify such a sub-dominant
operator -which he calls a hyperfermion-  with Wen's electron operator introduced in
the chiral LL\cite{stone}. This hyperfermion is identical to
the dual state for $K=1/(2n+1)$}. 
In general the dual state and the electron are
however orthogonal: this is quite clear when one considers the LL with spin.
The dual state is then generalized to a state with the same quantum numbers
as the electron (carrying a unit charge and a spin one-half), but with again
anyonic statistics. But due to spin-charge separation, that state is not
stable and decays into a spinless charge one quasiparticle which is none
other than the holon, and a spin one-half excitation, which is just the
spinon. The dual excitation we have found is therefore the analog of the
spinon and the holon for the spinless LL and has nothing to do (in general)
with an electron. In the following in accordance with the previous remarks
we will call these states holons (for the spinless LL) or dual states.

These dual holon states also occur in Haldane's interpretation of the
Calogero-Sutherland model: he proposed that a natural interpretation of such
a model was not in terms of electrons or bosons but as a gas of
non-interacting anyons\cite{hal94,zirn}. The basis for that interpretation
is the finding that for rational values of the coupling $\lambda =p/q$ ($%
\lambda $ is related to the LL parameter by the simple relation $\lambda
=1/K $), the dynamical structure factor obeys simple selection rules: $q$
''charge $-1$ bare particles'' (the anyon -it has anyonic statistics $\pi
\lambda =\pi /K$) are created with $p$ ''holes'' which therefore carry a
charge $1/\lambda $. The particles and holes appear as pseudo-particles in a
pseudo-momenta parametrization of the spectrum. There are however
difficulties with that interpretation: it is not clear how the
selection rules are generalized when the coupling $\lambda $ is irrational;
the physics is indeed completely continuous with the coupling while the
selection rules are only valid for rational couplings; besides, an
asymmetry is introduced between particles and holes. This means that this
parametrization which relies on pseudoparticles is probably inadequate. In
the low energy limit, the Calogero-Sutherland model has the properties of a
Luttinger liquid. It is therefore possible to describe its quasiparticle
spectrum in terms of the fractional excitations we have found in this paper:
we do find the charge $1$ anyon proposed (this is our dual state); however
our selection rules are quite different. First they depend on the
statistics (electrons or bosons): in contrast the pseudo-particle based
selection rules do not involve spinons; this runs contrary to results on the
bosonic LL for which the dual state must actually be seen as a composite
state made out of two spinons. Second, our selection rules are valid even
for irrational couplings {that is imply the existence of}
 quantum states with sharp irrational charges. Third
our selection rules respect particle-hole duality: there exist both a charge
one anyon with statistics $\pi /K$ (the dual holon) and a similar charge $-1$
anyon; the same applies to Laughlin quasiparticle for which we have both
quasiholes and quasielectrons. Our selection rules involve the hybrid state
and the Laughlin quasiparticles, and the holon only appears in the dual
basis where it is accompanied again by the hybrid quasiparticle.

Finally, a {symmetric basis} can be associated with the hybrid excitation: 
\begin{equation}
\left( 
\begin{tabular}{l}
$Q_{+}$ \\ 
$Q_{-}$%
\end{tabular}
\right) =m\left( 
\begin{tabular}{l}
$\frac{1+K}{2}$ \\ 
$\frac{1-K}{2}$%
\end{tabular}
\right) +n\left( 
\begin{tabular}{l}
$\frac{1-K}{2}$ \\ 
$\frac{1+K}{2}$%
\end{tabular}
\right) ,
\end{equation}
where $m=\frac{Q+J}{2}$ and $n=\frac{Q-J}{2}$; $m$ and $n$ are again
independent integers. They physically correspond to the number of electrons
added to the system at the right and left Fermi points respectively. That
self-dual basis reduce to Landau quasiparticles when $K=1$. The physical
processes { generated} in that symmetric basis are not charge or current
excitations but addition of electrons at the Fermi surface. Note that the
arbitrariness in the choice of a basis simply reflects the possibility to
stress various specific physical processes as elementary. But experiments
probe $Q_{+}$ and $Q_{-}$; for a given set of $Q$ and $J$, $Q_{\pm}$ assume
the same value irrespective of the basis choice.


\section{The Luttinger Liquid with spin.}

\label{ch-3}

In this section we generalize the construction of fractional excitations
developed in section (\ref{ch-two}) to the full Luttinger liquid with spin.
One of the main properties exhibited by the effective theory is 
spin-charge separation, the complete decoupling of spin and charge dynamics.
In the exact solution of the Hubbard model by Bethe Ansatz, excitations
display such a spin-charge separation: 
{one state, the holon is a spinless particle carrying the charge of the
electron, while the other, the spinon, is a neutral spin one-half state\cite{lieb}}.
This is an asymptotic property only valid in the low-energy limit (the Hubbard
model in the large $U$ limit is an exception because spin-charge separation
is realized at all energy scales). However this property is not {obtained 
for all the gapless itinerant 1D models, in the low energy limit}. According
to the universality hypothesis  they should be described by the
Luttinger liquid framework if the interaction is not too long-ranged. One
such example is the Hubbard model in a magnetic field which does not display
 spin-charge separation even in the low-energy limit, although it is a short
range gapless model. This model was analyzed  by Frahm and Korepin in
the framework of Bethe Ansatz plus conformal field theory\cite
{frahm}: they were  able to compute the anomalous exponents for the correlation
functions. Several issues remain unclear for such
models in a magnetic field: {in particular what the excitations are}. Since spin-charge
separation {does not occur}, the holon and the spinon cannot be  the
elementary excitations of the system anymore. {To  answer the question
we have to turn to the low-energy effective theory}. Frahm and Korepin's results imply that
an effective description in terms of the gaussian model should be possible
since conformal invariance is realized. We will find a generalization of the spin-charge
separated gaussian hamiltonian {suitable for a description of} the Hubbard model in a
magnetic field. {Our formalism is} very
similar to Wen's $K$ matrix approach to edge states of the FQHE. This will
enable us to characterize very precisely, in the low-energy limit, the
properties of 1D gapless models with or without spin-charge separation such
as the Hubbard model in a magnetic field: we will find that in the latter
case, although there is no spin-charge separation, there is still a
generalized decoupling. The excitations are again fractional; as expected
the holon and the spinon {are no longer present in} the spectrum and we will give
the general framework allowing the description of the fractional states
which replace them.

\subsection{Spin-charge separated Luttinger liquid.}

\label{ch-31}

We start with the standard case when spin-charge separation exists. 
Although fractional excitations are clearly present in Bethe
Ansatz, no description of these special states was attempted in the
low-energy limit through bosonization. {In the following we answer several questions:}
 how does the holon evolve with
interaction? What would be an effective wavefunction for it? Is it a semion?
First, we consider the
ground state of the two-component gaussian model, because it will suggest to us
a possible generalization of the gaussian model which will prove to be the
correct one for the description of gapless models without spin-charge
separation such as the Hubbard model in a magnetic field.

\subsubsection{Ground state of the gaussian hamiltonian.}

We consider a two component model by introducing an internal quantum number
such as the $SU(2)$ spin. We consider the charge and spin densities as well
as their associated phase fields:

\begin{eqnarray}
\rho _{c} &=&\rho _{\uparrow }+\rho _{\downarrow };\;\rho _{s}=\rho
_{\uparrow }-\rho _{\downarrow }, \\
\rho _{\sigma } &=&-\frac{1}{\sqrt{\pi }}\partial _{x}\Phi _{\sigma
},\;\sigma =\uparrow ,\downarrow \\
\left[ \Theta _{\sigma }(x),\partial _{x}\Phi _{\sigma ^{\prime }}(y)\right]
&=&i\delta _{\sigma \sigma ^{\prime }}\delta (x-y), \\
\Phi _{c/s} &=&\frac{\Phi _{\uparrow }\pm \Phi _{\downarrow }}{\sqrt{2}}, \\
\left[ \Theta _{\tau }(x),\partial _{x}\Phi _{\tau ^{\prime }}(y)\right]
&=&i\delta _{\tau \tau ^{\prime }}\delta (x-y);\;\tau =c,s.
\end{eqnarray}
The effective hamiltonian derived for instance from the Hubbard model in the
absence of a magnetic field is:

\begin{eqnarray}
H &=&H_{c}+H_{s}, \\
H_{\tau } &=&\frac{u_{\tau }}{2}\int_{0}^{L}dx\ K_{\tau }^{-1}\left( \nabla
\Phi _{\tau }\right) ^{2}+K_{\tau }(\nabla \Theta _{\tau })^{2};\;\tau =c,s.
\end{eqnarray}
$\nabla \Phi _{\tau }=\Pi _{\tau }$ is canonically conjugate to the field $%
\Theta _{\tau }$. One easily extracts the ground state which is simply a
product of gaussians: 
\begin{equation}
\Psi _{0}(\{\Pi _{\tau ,q_{n}}\})=\prod_{\tau =c,s}\exp (-\frac{1}{2K_{\tau }%
}\sum_{n\neq 0}\frac{1}{\left| q_{n}\right| }\Pi _{\tau ,q_{n}}\Pi _{\tau
,-q_{n}}).
\end{equation}
In terms of charge and spin densities: 
\begin{equation}
\Psi _{0}(\{\rho _{\tau ,q_{n}}\})=\prod_{\tau =c,s}\exp (-\frac{1}{4K_{\tau
}}\sum_{n\neq 0}\frac{\pi }{\left| q_{n}\right| }\rho _{\tau ,q_{n}}\rho
_{\tau ,-q_{n}}).
\end{equation}
The ground state displays of course a complete decoupling of spin and charge
as is apparent from the previous expression. This is also a Jastrow
wavefunction. In real space: 
\[
\Psi _{0}(\rho _{\tau })=\prod_{\tau =c,s}\exp \frac{1}{4K_{\tau }}[ 
\]
\begin{equation}
\int \int dxdx^{\prime }\rho _{\tau }(x)\ln \left| \sin \frac{\pi
(x-x^{\prime })}{L}\right| \rho _{\tau }(x^{\prime })].
\end{equation}
We define the charge and spin parts of the ground state per: 
\begin{eqnarray}
\Psi _{c/s} &=&\exp \frac{1}{4K_{c/s}}[  \nonumber \\
&&\int \int dxdx^{\prime }\rho _{c/s}(x)\ln \left| \sin \frac{\pi
(x-x^{\prime })}{L}\right| \rho _{c/s}(x^{\prime })].  \label{chargespin}
\end{eqnarray}
The previous ground state may be rewritten in terms of the densities of each
species: 
\[
\Psi _{0}(\{\rho _{\sigma }\})=\exp \frac{1}{2}[ 
\]
\begin{equation}
\int \int dxdx^{\prime }\rho _{\sigma }(x)g_{\sigma \sigma ^{\prime }}\ln
\left| \sin \frac{\pi (x-x^{\prime })}{L}\right| \rho _{\sigma ^{\prime
}}(x^{\prime })],  \label{plasma}
\end{equation}
where we have introduced the following $\widehat{g}$ matrix: 
\begin{equation}
g_{\sigma \sigma ^{\prime }}=\left( 
\begin{tabular}{ll}
$\frac{K_{c}^{-1}+K_{s}^{-1}}{2}$ & $\frac{K_{c}^{-1}-K_{s}^{-1}}{2}$ \\ 
$\frac{K_{c}^{-1}-K_{s}^{-1}}{2}$ & $\frac{K_{c}^{-1}+K_{s}^{-1}}{2}$%
\end{tabular}
\right)
\end{equation}
The eigenvalues of that matrix are simply the inverses of the Luttinger
liquid parameters $K_{c}^{-1}$ and $K_{s}^{-1}$. If one rewrites the
wavefunction in terms of individual electron coordinates $\rho _{\sigma
}(x)=\sum_{i,\sigma }\delta (x-x_{i})$ (the sum is restricted to particles
with spin $\sigma $) and if one sets $z=\exp i\frac{2\pi }{L}x$, one easily
finds: 
\begin{equation}\label{multiboson}
\Psi _{0}(\{x_{i},\sigma _{i}\})=\prod_{i<j}\left| z_{i}-z_{j}\right|
^{g_{\sigma _{i}\sigma _{j}}}.
\end{equation}
The wavefunction is bosonic; for the fermionic LL, one undoes the
Jordan-Wigner transformation which leads to: 
\[
\Psi _{F,0}(\{x_{i},\sigma _{i}\})=\prod_{i<j}\left| z_{i}-z_{j}\right|
^{g_{\sigma _{i}\sigma _{j}}} 
\]
\begin{equation}\label{multifermion}
\times \prod_{i<j}\left\{ \left( \frac{(z_{i}-z_{j})}{\left|
z_{i}-z_{j}\right| }\right) ^{\delta _{\sigma _{i}\sigma _{j}}}\exp i\frac{%
\pi }{2}sgn(\sigma _{i}-\sigma _{j})\right\} .
\end{equation}
(The antisymmetrizing factor {consists} of two parts, one which ensures
{that particles of the same species anticommute}, and a second part known
as a Klein factor which allows antisymmetry for particles of different
spin.) Let us redefine the matrix elements of $\widehat{g}$ per:

\begin{eqnarray}
g_{\sigma \sigma ^{\prime }} &=&\left( 
\begin{tabular}{ll}
$\lambda $ & $\mu $ \\ 
$\mu $ & $\lambda $%
\end{tabular}
\right) \\
K_{c}^{-1} &=&\lambda +\mu \\
K_{s}^{-1} &=&\lambda -\mu
\end{eqnarray}
If we denote the coordinates of particles with spin $\uparrow $ and $%
\downarrow $ respectively by $u$ and $v$ then the ground state can be rewritten
as (for convenience the antisymmetrizing factor is omitted):

\begin{equation}
\Psi _{0}(\{u_{i},v_{i}\})=\prod_{i<j}\left| u_{i}-u_{j}\right| ^{\lambda
}\prod_{i<j}\left| v_{i}-v_{j}\right| ^{\lambda }\prod_{i,j}\left|
u_{i}-v_{j}\right| ^{\mu },
\end{equation}

\begin{eqnarray}
\Psi _{c} &=&\left[ \prod_{i<j}\left| u_{i}-u_{j}\right| \left|
v_{i}-v_{j}\right| \left| u_{i}-v_{j}\right| \right] ^{1/2K_{c}}  \nonumber
\\
&=&\prod_{i<j}\left| z_{i}-z_{j}\right| ^{1/2K_{c}},  \label{psi-c}
\end{eqnarray}
\begin{eqnarray}
\Psi _{s} &=&\left[ \prod_{i<j}\left| u_{i}-u_{j}\right| \left|
v_{i}-v_{j}\right| /\left| u_{i}-v_{j}\right| \right] ^{1/2K_{s}}  \nonumber
\\
&=&\prod_{i<j}\left| z_{i}-z_{j}\right| ^{\sigma _{i}\sigma _{j}/2K_{s}}.
\label{psi-s}
\end{eqnarray}

For the fermionic LL the charge part gets an additional factor $%
\left( \frac{(x_{i}-x_{j})}{\left| x_{i}-x_{j}\right| }\right) ^{1/2}$ and 
the spin part, a factor $%
\left( \frac{(x_{i}-x_{j})}{\left| x_{i}-x_{j}\right| }\right) ^{\sigma
_{i}\sigma _{j}/2}\exp i\frac{\pi }{2}sgn(\sigma _{i}-\sigma _{j})$.
 
 These are 1D Laughlin multi-component wavefunctions. In 2D they are known
as Halperin wavefunctions which describe multi-component
systems of the FQHE\cite{halperin}. In that context the $\widehat{g}$ matrix
is known as Wen's topological $K$ matrix\cite{wen}. The main difference
between the two matrices is that the the entries of the $K$ matrix 
are integers while $%
\widehat{g}$ matrix elements are arbitrary real numbers ({only constrained 
to yield positive and real eigenvalues}). The $\widehat{g}$
matrix does not allow for a topological interpretation either since there is
no topological quantization as in the FQHE. We will call the $\widehat{g}$
matrix, the {\it charge matrix} because it corresponds to the couplings
between particles in the plasma analogy (see eq.(\ref{plasma}))\cite{f14}.


\subsubsection{Elementary excitations : the holon, the spinon, Laughlin
quasiparticles.}

We generalize the approach followed for the spinless LL. There is a
decoupling of the dynamics at two levels: chiral separation as well as
spin-charge separation. In particular both the charge and the spin
hamiltonians - $H_{c}$ and $H_{s}$ - display chiral separation: $%
H_{c}=H_{c+}+H_{c-}$ and $H_{s}=H_{s+}+H_{s-}$ where the four hamiltonians
all commute ($\left[ H_{c/s\pm },H_{c/s\pm }\right] =0$). The
following operators create the exact eigenstates of the relevant chiral
hamiltonians: 
\begin{equation}
V_{\tau }^{\pm }(Q_{\tau ,\pm },q)=\int dx\exp iqx\exp -i\sqrt{\pi /2}%
Q_{\tau ,\pm }\Theta _{\tau ,\pm }
\end{equation}
\begin{eqnarray}
\Theta _{\tau ,\pm } &=&\Theta _{\tau }\mp \Phi _{\tau }/K_{\tau };\;\tau
=c,s, \\
q &=&\frac{2\pi n}{L}\mp \frac{2\pi }{L}\frac{Q_{\tau ,\pm }^{2}}{K_{\tau }},
\\
Q_{\tau ,\pm } &=&\frac{Q_{\uparrow }+\tau Q_{\downarrow }}{2}\pm K_{\tau }%
\frac{J_{\uparrow }+\tau J_{\downarrow }}{2}.
\end{eqnarray}
(The square root $\sqrt{2}$ in the exponential comes from the normalization
of the charge and spin fields; $c$ and $s$ index charge and spin
respectively; $\tau =\pm 1$ for charge and spin respectively.) One 
easily checks
that: 
\begin{equation}
\left[ \widehat{Q}_{\uparrow }+\tau \widehat{Q}_{\downarrow },V^{\pm
}(Q_{\tau ^{\prime },\pm })\right] =\delta _{\tau \tau ^{\prime }}Q_{\tau
,\pm }V^{\pm }(Q_{\tau ^{\prime },\pm }).
\end{equation}
This implies that these excitations either carry a charge $Q=Q_{c,\pm }$ but
then have no spin (the operators $V_{c}^{\pm }$), or that they have a spin $%
S_{z}=S_{\pm }=Q_{s,\pm }/2$ but no charge (operators $V_{s}^{\pm }$).As
expected the fractional states come in two brands: the first corresponds to
charge excitations and the second to spin excitations. Hereafter we will
note the charge and the spins of these excitations as

\begin{eqnarray}
Q_{\pm } &=&Q_{c\pm }, \\
S_{\pm } &=&\frac{Q_{s,\pm }}{2}.
\end{eqnarray}
Because of the obvious relevance to physical systems we focus first on
elementary excitations for a fermionic LL; we will consider the case of a
bosonic LL later in section (\ref{ch-33}). Solving the constraints on the
charge and current which again obey the selection rule $Q_{\uparrow
}-J_{\uparrow }=2n_{\uparrow }$ and $Q_{\downarrow }-J_{\downarrow
}=2n_{\downarrow }$ ( the $n$ are integers), we find: 
\begin{eqnarray}
\left( 
\begin{tabular}{l}
$Q_{+}$ \\ 
$Q_{-}$ \\ 
$S_{+}$ \\ 
$S_{-}$%
\end{tabular}
\right) &=&n_{\uparrow }\left( 
\begin{tabular}{l}
$1$ \\ 
$1$ \\ 
$1/2$ \\ 
$1/2$%
\end{tabular}
\right) +n_{\downarrow }\left( 
\begin{tabular}{l}
$1$ \\ 
$1$ \\ 
$-1/2$ \\ 
$-1/2$%
\end{tabular}
\right)  \nonumber \\
&&+J_{\uparrow }\left( 
\begin{tabular}{l}
$\frac{1+K_{c}}{2}$ \\ 
$\frac{1-K_{c}}{2}$ \\ 
$\frac{1+K_{s}}{4}$ \\ 
$\frac{1-K_{s}}{4}$%
\end{tabular}
\right) +J_{\downarrow }\left( 
\begin{tabular}{l}
$\frac{1+K_{c}}{2}$ \\ 
$\frac{1-K_{c}}{2}$ \\ 
$-\frac{1+K_{s}}{4}$ \\ 
$-\frac{1-K_{s}}{4}$%
\end{tabular}
\right)  \label{states}
\end{eqnarray}
This compact equation must be read as follows. Each entry represents a
fractional excitation; the first two lines are charge spinless excitations,
while the last two lines represent spin excitations. For instance the entry $%
Q_{+}$ is associated with a fractional excitation with charge $Q=Q_{+}$
which carries no spin, and propagates in the right direction: therefore $%
\frac{1+K_{c}}{2}$ in the first line means a spinless state with charge $Q=%
\frac{1+K_{c}}{2}$ going to the right. Likewise the second line
characterizes charge excitations propagating to the left. The line $S_{+}$
means that the states have no charge, a spin component $S_{z}=S_{+}$ and
propagate to the right: for instance $1/2$ is a spin one-half fractional
state. Each line gives the decomposition of a given fractional excitation
into elementary excitations: for instance the $Q_{+}$ excitation is made up
of $n_{\uparrow }+n_{\downarrow }$ excitations $V_{c}^{+}(Q=1)$, and $%
J_{\uparrow }+J_{\downarrow }$ excitations $V_{c}^{+}(Q=\frac{1+K_{c}}{2})$.
The previous equation summarizes the selection rules which are obeyed by the
elementary excitations.

Let us give an example. Suppose one adds a spin up electron at the Fermi
level in the Luttinger liquid. This is a $Q_{\uparrow }=1=J_{\uparrow }$ and 
$Q_{\downarrow }=0=J_{\downarrow }$ excitation or in terms of $n_{\uparrow }$
and $n_{\downarrow }$, this is a ($n_{\uparrow }=0$, $n_{\downarrow }=0,$ $%
J_{\uparrow }=1,$ $J_{\downarrow }=0$) state. Equation (%
\ref{states}) shows that the spinor 
\begin{equation}
\left( 
\begin{tabular}{l}
$Q_{+}$ \\ 
$Q_{-}$ \\ 
$S_{+}$ \\ 
$S_{-}$%
\end{tabular}
\right) =\left( 
\begin{tabular}{l}
$\frac{1+K_{c}}{2}$ \\ 
$\frac{1-K_{c}}{2}$ \\ 
$\frac{1+K_{s}}{4}$ \\ 
$\frac{1-K_{s}}{4}$%
\end{tabular}
\right)
\end{equation}
is created; this means that the spin up electron added at the Fermi level $%
k_{F}$ splits into four fractional states: a charge $\frac{1+K_{c}}{2%
}$ anyon propagating at velocity $u_{c}$; this state has no spin; 
a second charge anyon with charge $\frac{1-K_{c}}{2}$ and velocity $-u_{c}$
and then two spin anyons with velocities $u_{s}$ and $-u_{s}$ and
respective spin $S_{z}=\frac{1\pm K_{s}}{4}$. In the special case of spin
rotational invariance ($K_{s}=1$) there is only one spin anyon: the spinon
with spin $S_{z}=1/2$ which propagates to the right with velocity $u_{s}$
( or to the left with velocity $-u_{s}$ if the electron had been added at the
left Fermi point $-k_{F}$). Likewise, if $K_{c}=1$, there is a single charge
state, which has charge $Q=1$. In the non-interacting case, the charge
velocity $u_{c}$ and the spin velocity $u_{s}$ are equal and therefore the
spinon and the charge $1$ state subsume into a single state since they move
in the same direction with the same velocity: we have just recovered the
spin up electron.

\paragraph{Holon and spinon.}

Let us identify the content of the elementary excitations, starting with $%
V_{c}^{\pm }(Q=1)$ and $V_{s}^{\pm }(S_{z}=1/2)$. The charge and the spin
carried by these fractional excitations make it reasonable to interpret them
as the holon and the spinon respectively.

The physical processe involved in the creation of 
each of  these states 
confirms this identification. Indeed the minimal operation which involves $%
V_{c}^{\pm }(Q=1)$ is obtained when $J_{\uparrow }=0=J_{\downarrow }$ and
set $n_{\uparrow }=1=n_{\downarrow }$ in equation (\ref{states}). This is an
excitation for which $Q_{\uparrow }=1=Q_{\downarrow }$ and $J_{\uparrow
}=0=J_{\downarrow }$ which means that this is a pure charge process (no spin
variation $S_{z}=(Q_{\uparrow }-Q_{\downarrow })/2=0$, no spin current nor
charge current). $V_{c}^{\pm }(Q=1)$ is an excitation associated with the
addition of charge in the LL. All {the transitions in Fock space
which occur after adding a charge to the ground state 
therefore involve 
$V_{c}^{\pm }(Q=1).$ It is then consistent to identify $V_{c}^{\pm }(Q=1)$ as
the holon\cite{holon-charge}}.

Likewise the minimal excitation generating $V_{s}^{\pm }(S_{z}=1/2)$ is a
spin one transition which is a pure spin process. All excitations for which
there is a spin flip will therefore create $V_{s}^{\pm }(S_{z}=1/2)$ (in
pairs). This is what we expect from a spinon.

Notice that both for the holon and spinon there are even-odd effects arising
in the low-energy gaussian theory. Indeed, equation (\ref{states}) shows that
 an excitation with includes a spin one-half transition will not
create a spinon (we have instead a ''hybrid'' spin excitation): one needs at
least a spin one transition to crate a
spinon. This means that adding a single electron does not
create a spinon: 
{even number of electrons are required}. {This makes sense } since for a spin chain the
minimal spin excitation is also a spin-flip which involves two electrons and
not just one. The same behaviour is observed with the holon: the minimal
process which creates it {adds two electrons} ($Q=2$ since $%
Q_{\uparrow }=1=Q_{\downarrow }$). 
These
even-odd effects are a direct consequence of statistics and would not be
observed with a bosonic two component LL.

With eq.(\ref{chargespin}), we find that the wavefunctions for the holon and
the spinon are simply (with $z=\exp i\frac{2\pi }{L}x$): 
\begin{eqnarray}
\Psi _{holon}(z_{0}) &=&V_{c}^{+}(Q_{c}=1,z_{0})\Psi _{c}  \nonumber \\
&=&\prod_{i}\left( \frac{(x_{i}-x_{0})}{\left| x_{i}-x_{0}\right| }\right) ^{%
\frac{1}{2K_{c}}}\prod_{i}\left| z_{i}-z_{0}\right| ^{1/2K_{c}}  \nonumber
\end{eqnarray}

\begin{eqnarray}\label{hoho}
&&\times \prod_{i<j}\left| z_{i}-z_{j}\right|
^{1/2K_{c}}\prod_{x_{0},i<j}\left\{ \frac{(x_{i}-x_{j})}{\left|
x_{i}-x_{j}\right| }\right\} ^{1/2} \\
&=&\prod_{i}\left( z_{i}-z_{0}\right) ^{1/2K_{c}}\prod_{i<j}\left|
z_{i}-z_{j}\right| ^{1/2K_{c}}  \nonumber \\
&&\times \exp -i\frac{2k_{F}}{K_{c}}\left( \frac{\sum_{i}x_{i}}{N}%
+x_{0}\right) \prod_{x_{0},i<j}\left\{ \frac{(x_{i}-x_{j})}{\left|
x_{i}-x_{j}\right| }\right\} ^{1/2},
\end{eqnarray}
and: 
\begin{eqnarray}\label{spispi}
&&\Psi _{spinon}(\sigma _{0},z_{0})  \nonumber \\
&=&V_{s}^{+}(S_{z}=\sigma _{0}/2,z_{0})\Psi _{s}  \nonumber \\
&=&\prod_{i}\left( \frac{(x_{i}-x_{0})}{\left| x_{i}-x_{0}\right| }\right)
^{\sigma _{0}\sigma _{i}/2K_{s}}\prod_{i}\left| z_{i}-z_{0}\right| ^{\sigma
_{0}\sigma _{i}/2K_{s}}  \nonumber \\
&&\times \prod_{i<j}\left| z_{i}-z_{j}\right| ^{\sigma _{i}\sigma
_{j}/2K_{s}}\prod_{x_{0},i<j}\left\{ \frac{(x_{i}-x_{j})}{\left|
x_{i}-x_{j}\right| }\right\} ^{\sigma _{i}\sigma _{j}/2}.
\end{eqnarray}

The holon and the spinon are both anyons with exchange statistics:

\begin{eqnarray}
\theta _{c} &=&\frac{\pi }{2K_{c}}, \\
\theta _{s} &=&\frac{\pi }{2K_{s}}
\end{eqnarray}
(the statistics were computed in subsection \ref{sub1-2}). {Except for the
special case $K_{\tau}=1$, these objects are not 
semions}; 
in addition, for holons we must also require 
$v_{c}\neq
v_{s}$ to ensure spin-charge
separation. {Contrast our results (eqns (\ref{hoho}) and 
(\ref{spispi})) with the commonly used but} incorrect
charge-spin decoupling of the electron operator\cite{f15}. The holon and the
spinon generalize the dual excitation found for the spinless LL; in the same
way, duality transforms the holon and the spinon into the
two-component generalizations of the Laughlin quasiparticles. This is a most
remarkable yet simple result because it shows that two seemingly
unrelated fractional excitations -the holon (or spinon) and the Laughlin
quasiparticle- occurring in two very different contexts are actually deeply
connected.

Similarly to the spinless LL, in addition to the holon and to the spinon, we
hybrid excitations complete the basis of fractional
excitations. Their charge and spin are intermediate between those of the
holon and spinon and those of their dual excitations, the Laughlin
quasiparticles which we discuss now.

\paragraph{Laughlin quasiparticles.}

We can choose another basis of elementary excitations dual to the previous
one which will parametrize the excitations in terms of current processes and
electron addition at the Fermi surface. This basis, emphasizing Laughlin
quasiparticles as elementary excitations reads: 
\begin{eqnarray}
\left( 
\begin{tabular}{l}
$Q_{+}$ \\ 
$Q_{-}$ \\ 
$S_{+}$ \\ 
$S_{-}$%
\end{tabular}
\right) &=&n_{\uparrow }\left( 
\begin{tabular}{l}
$K_{c}$ \\ 
$-K_{c}$ \\ 
$K_{s}/2$ \\ 
$-K_{s}/2$%
\end{tabular}
\right) +n_{\downarrow }\left( 
\begin{tabular}{l}
$K_{c}$ \\ 
$-K_{c}$ \\ 
$-K_{s}/2$ \\ 
$K_{s}/2$%
\end{tabular}
\right)  \nonumber \\
&&+Q_{\uparrow }\left( 
\begin{tabular}{l}
$\frac{1+K_{c}}{2}$ \\ 
$\frac{1-K_{c}}{2}$ \\ 
$\frac{1+K_{s}}{4}$ \\ 
$\frac{1-K_{s}}{4}$%
\end{tabular}
\right) +Q_{\downarrow }\left( 
\begin{tabular}{l}
$\frac{1+K_{c}}{2}$ \\ 
$\frac{1-K_{c}}{2}$ \\ 
$-\frac{1+K_{s}}{4}$ \\ 
$-\frac{1-K_{s}}{4}$%
\end{tabular}
\right) ,  \label{excite}
\end{eqnarray}
where again $Q_{\uparrow }-J_{\uparrow }=2n_{\uparrow }$ and $Q_{\downarrow
}-J_{\downarrow }=2n_{\downarrow }.$ In addition to the hybrid
quasiparticles $V_{c}^{\pm }(Q=\frac{1\pm K_{c}}{2})$ and $V_{s}^{\pm
}(S_{z}=\frac{1\pm K_{s}}{4})$ which already existed in the previous basis,
we have two excitations associated with pure charge current or spin current
processes: $V_{c}(Q_{c}=K_{c})${\it \ }and{\it \ }$V_{s}(S_{z}=K_{s}/2)${\it %
. }Under electromagnetic duality the latter are conjugate to the holon and
spinon respectively. Actually they are obtained by spin-charge separation of
the two-component Laughlin quasiparticle{\it ,} and we may call them a
Laughlin holon and a Laughlin spinon. Let us compute the
wavefunctions of these two excitations; one finds: 
\begin{eqnarray}\label{creho}
&&V_{c}(Q_{c} =K_{c},z_{0})\Psi _{c}  \nonumber \\
&=&\prod_{i}(z_{i}-z_{0})^{1/2} \nonumber\\
&&\times \exp -i2k_{F}\left( \frac{\sum x_{i}}{N}+x_{0}\right) \Psi _{c},
\end{eqnarray}
\begin{eqnarray}
&&V_{s}(S^{z}=\sigma _{0}K_{s}/2,z_{0})\Psi _{s} \nonumber \\
&=&\prod_{i}(z_{i}-z_{0})^{\sigma _{i}\sigma _{0}/2} \nonumber \\
&\times & \exp -i2\sigma _{0}(k_{\uparrow }-k_{\downarrow })\left( \frac{\sum
x_{i\uparrow }-x_{j\downarrow }}{M}+x_{0}\right) \Psi _{s} \\
\label{crespi}
\end{eqnarray}

In the previous expression $\sigma _{0}$ takes on the values $\pm 1$ and $%
k_{\uparrow },k_{\downarrow }$ are the Fermi vectors associated with
particles of spin up and down: $k_{\sigma }=\frac{\pi }{L}N_{\sigma }$ and $%
N=N_{\uparrow }-N_{\downarrow }$, $M=N_{\uparrow }-N_{\downarrow }$. $\Psi
_{c}$ and $\Psi _{s}$ are given for the bosonic LL in eq.(\ref{psi-c}) and (%
\ref{psi-s}); for the fermionic LL, there are additional phase factors given
in the text following eq.(\ref{psi-c}) and (\ref{psi-s}). Plasma analogy 
allows to get the charge and spin of the Laughlin
holon eq.(\ref{creho}) and spinon eq.(\ref{crespi})).
{The Laughlin holon and spinon are the stable excitations into which the
Laughlin quasiparticle decays as a 
result of spin-charge separation};
{in the localized Wannier
basis we have considered throughout,  the product of the
wavefunctions of the two excitations yields indeed}: 
\begin{equation}
\Psi _{qp}(z_{0},\sigma _{0})=\prod_{i}(z_{i}-z_{0})^{\delta _{\sigma
_{0}\sigma _{i}}}\prod_{i<j}\left| z_{i}-z_{j}\right| ^{g_{\sigma _{i}\sigma
_{j}}},
\end{equation}
which is just the generalization of the Laughlin quasiparticle to two
component systems: when we add spin, the Laughlin quasiparticle comes in
two flavours (up or down) and the Laughlin correlation hole acts only on
particles of the same flavour. It therefore carries both fractional charge
and fractional spin. (It would be spinless if the Laughlin prefactor were $%
\prod_{i}(z_{i}-z_{0})$ instead of $\prod_{i}(z_{i}-z_{0})^{\delta _{\sigma
_{0}\sigma _{i}}}$).

The Laughlin holon and spinon have statistical phases $\theta _{c}=\pi K_{c}$
and $\theta _{s}=\pi K_{s}/2$. For a spin-rotational invariant system, the
Laughlin spinon and its dual conjugate - the spinon - are identical states ($%
K_{s}=1$ is the self-dual point for spin excitations).


\subsection{Luttinger liquid without spin-charge separation.}

\label{ch-32}

\subsubsection{The general LL and the charge matrix.}

We now generalize the standard LL theory to include situations with no
spin-charge separation. We start from the ground state of the gaussian
hamiltonian: 
\begin{equation}
\Psi _{0}\left[ \widehat{g}\right] =\exp \frac{1}{2}\int \int dxdx^{\prime
}\rho _{\sigma }(x)g_{\sigma \sigma ^{\prime }}\ln \left| \sin \frac{\pi
(x-x^{\prime })}{L}\right| \rho _{\sigma ^{\prime }}(x^{\prime }),
\end{equation}

\begin{eqnarray}
g_{\sigma \sigma ^{\prime }} &=&\left( 
\begin{tabular}{ll}
$\lambda $ & $\mu $ \\ 
$\mu $ & $\lambda $%
\end{tabular}
\right) , \\
K_{c}^{-1} &=&\lambda +\mu , \\
K_{s}^{-1} &=&\lambda -\mu .
\end{eqnarray}
and relax the constraint $g_{\uparrow \uparrow }=g_{\downarrow \downarrow
}$ (while $\widehat{g}$ is kept symmetric). We consider the charge matrix: $%
g_{\sigma \sigma ^{\prime }}=\left( 
\begin{tabular}{ll}
$\lambda $ & $\mu $ \\ 
$\mu $ & $\lambda ^{\prime }$%
\end{tabular}
\right) $ and the associated wavefunction $\Psi _{0}\left[ \widehat{g}%
\right] $. We introduce for convenience the eigenvalues of the charge matrix
and the unitary matrix $P$ :

\begin{eqnarray}
P^{-1}\widehat{g}P &=&\widehat{D}=\left( 
\begin{tabular}{ll}
$1/K_{1}$ & $0$ \\ 
$0$ & $1/K_{2}$%
\end{tabular}
\right) \\\label{changebaz}
P_{\sigma \tau }P_{\sigma ^{\prime }\tau ^{\prime }}g_{\sigma \sigma
^{\prime }} &=&\frac{\delta _{\tau \tau ^{\prime }}}{K_{\tau }}\;;\quad \tau
=1,2
\end{eqnarray}
The normal modes of the charge matrix are simply: 
\begin{equation}\label{dansite}
\rho _{\tau }=P_{\sigma \tau }\rho _{\sigma }\Leftrightarrow \rho _{\sigma
}=P_{\sigma \tau }\rho _{\tau }\;\tau =1,2
\end{equation}
\begin{equation}
\sum_{\sigma \sigma ^{\prime }}\rho _{\sigma }(x)g_{\sigma \sigma ^{\prime
}}\rho _{\sigma ^{\prime }}(x^{\prime })=\sum_{\tau }\rho _{\tau }(x)\frac{1%
}{K_{\tau }}\rho _{\tau }(x^{\prime }).
\end{equation}
If the charge matrix obeys a $Z_{2}$ symmetry then the normal modes are just
the charge and spin density (up to a normalization factor) $\rho _{1}=\rho
_{c}/\sqrt{2}$ and $\rho _{2}=\rho _{s}/\sqrt{2}$). The wavefunction 
{now reads}: 
\begin{eqnarray}
&&\Psi _{0}[\widehat{g}]  \nonumber \\
&=&\exp \frac{1}{2K_{1}}\int \int dxdx^{\prime }\rho _{1}(x)\ln \left| \sin 
\frac{\pi (x-x^{\prime })}{L}\right| \rho _{1}(x^{\prime })  \nonumber
\end{eqnarray}
\begin{equation}
\times \exp \frac{1}{2K_{2}}\int \int dxdx^{\prime }\rho _{2}(x)\ln \left|
\sin \frac{\pi (x-x^{\prime })}{L}\right| \rho _{2}(x^{\prime }).
\end{equation}

We have expressed the ground state in this decoupled form because this allows us
to directly write down a gaussian hamiltonian with  ground state $\Psi _{0}[%
\widehat{g}]$. This generalizes the spin-charge decoupled gaussian
theory. We introduce the phase fields associated with the normal densities $%
\rho _{1}$ and $\rho _{2}$%
\begin{equation}\label{dansitebis}
\rho _{\tau }=-\frac{1}{\sqrt{\pi }}\partial _{x}\Phi _{\tau },\;j_{\tau }=%
\frac{1}{\sqrt{\pi }}\partial _{x}\Theta _{\tau },
\end{equation}
\begin{equation}
\left[ \Phi _{\tau }(x),\partial _{x}\Theta _{\tau ^{\prime }}(y)\right]
=i\delta _{\tau \tau ^{\prime }}\delta (x-y).
\end{equation}
It is then clear that $\Psi _{0}[\widehat{g}]$ is the exact ground state of
the following family of two component gaussian hamiltonians for arbitrary
velocities $u_{1},u_{2}$: 
\begin{eqnarray}\label{hache}
&&H[\widehat{g},u_{1},u_{2}]  \nonumber \\
&=&H_{B}[u_{1},K_{1}]+H_{B}[u_{2},K_{2}]  \nonumber \\
&=&\sum_{\tau =1,2}\frac{u_{\tau }}{2}\int_{0}^{L}dx\left[ K_{\tau
}^{-1}\left( \partial _{x}\Phi _{\tau }\right) ^{2}+K_{\tau }\left( \partial
_{x}\Theta _{\tau }\right) ^{2}\right] .
\end{eqnarray}
The two hamiltonians $H_{B}[u_{1},K_{1}]$ and $H_{B}[u_{2},K_{2}]$ commute
by construction. The next section will be devoted to the properties of that
generalized LL theory. {Let us stress here} the main property of this
general LL now: by construction that theory corresponds to {\it a
generalized separation}: the normal modes will not be charge and spin modes
but mix charge and spin in a proportion fixed in time. This will translate
for the fractional excitations to states with both fractional charge and
fractional spin.


\subsubsection{Main properties.}

The compressibility and spin susceptibility are easily computed and one
finds: 
\begin{equation}
\kappa ^{-1}=\frac{1}{L}\frac{\partial ^{2}E_{0}}{\partial \rho _{0}^{2}}=%
\frac{\pi }{4}\sum_{\tau }\frac{u_{\tau }}{K_{\tau }}\left( \sum_{\sigma
}P_{\sigma \tau }\right) ^{2},
\end{equation}

\begin{equation}
\chi _{s}^{-1}=\frac{1}{L}\frac{\partial ^{2}E_{0}}{\partial \rho _{s}^{2}}=%
\frac{\pi }{4}\sum_{\tau }\frac{u_{\tau }}{K_{\tau }}\left( \sum_{\sigma
}\sigma P_{\sigma \tau }\right) ^{2}.
\end{equation}

$\rho _{0}$ and $\rho _{s}$ are the charge and spin mean densities, while $%
E_{0}$ is the ground state energy. The Drude peak is: 
\begin{equation}
D=\frac{1}{L}\frac{\partial ^{2}E_{0}}{\partial \phi ^{2}}=\sum_{\tau
}u_{\tau }K_{\tau }\left( \sum_{\sigma }P_{\sigma \tau }\right) ^{2},
\end{equation}
where $\phi $ is a flux threading the LL ring of length $L$. This expression
can also be recovered using the Kubo formula; one then needs the
expression of the current density which one finds with the continuity
equation. The current is renormalized as for the spinless LL and the LL
with spin-charge separation. 
{In contrast to the case of the LL with spin-charge separation, here}
the expression involves both $K_{1}$ and $K_{2}$
because both modes one and two involve charge: 
\begin{equation}
j_{R}(x)=\sum_{\sigma }\left( \sum_{\sigma ^{\prime },\tau }u_{\tau }K_{\tau
}P_{\sigma \tau }P_{\sigma ^{\prime }\tau }\right) \frac{\partial _{x}\Theta
_{\sigma }}{\sqrt{\pi }}.
\end{equation}

Anomalous exponents are easily computed as functions of the charge
matrix $\widehat{g}$ which leads to compact expressions valid both for the
LL with spin-charge separation or for the more general LL; one introduces
the Fermi vectors: $k_{F\sigma }=\frac{\pi N_{\sigma }}{L}.$ {For instance
density-density correlators are
obtained with $H[\widehat{g}]$ and with the bosonization formulas,
yielding the static structure factor}; the
dominant Fourier components are $k=0$, $k=2k_{F\uparrow }$, $%
k=2k_{F\downarrow }$ and $k=2k_{F\uparrow }+2k_{F\downarrow }$. Near 
$k=0$:

\begin{eqnarray}
&<&\delta \rho _{\sigma }(0)\delta \rho _{\sigma ^{\prime }}(x)>_{k=0}=\frac{%
\widehat{g^{-1}}_{\sigma \sigma ^{\prime }}}{2(\pi x)^{2}}, \\
&<&\delta \rho (0)\delta \rho (x)>=\frac{A_{k=0}}{(\pi x)^{2}}%
;\;A_{k=0}=\sum_{\sigma \sigma ^{\prime }}\frac{\widehat{g^{-1}}_{\sigma
\sigma ^{\prime }}}{2}.
\end{eqnarray}

For the higher harmonics one includes a mode at $2k_{F\uparrow
}+2k_{F\downarrow }=2\pi \rho $ (which appears in the Hubbard model in a
magnetic field):

\begin{eqnarray}
&<&\delta \rho (0)\delta \rho (x)>=\frac{A_{k=0}}{(\pi x)^{2}}+a_{\uparrow }%
\frac{\cos (2k_{F\uparrow }x)}{x^{2+\alpha (2k_{F\uparrow })}}+a_{\downarrow
}\frac{\cos 2k_{F\downarrow }x}{x^{2+\alpha (2k_{F\downarrow })}}  \nonumber
\\
&&+b\frac{\cos (2k_{F\uparrow }+2k_{F\downarrow })x}{x^{2+\alpha
(2k_{F\uparrow }+2k_{F\downarrow })}},
\end{eqnarray}

\begin{eqnarray}
2+\alpha (2k_{F\sigma }) &=&2\widehat{g}_{\sigma \sigma }^{-1}, \\
2+\alpha (2k_{F\uparrow }+2k_{F\downarrow }) &=&2\sum_{\sigma \sigma
^{\prime }}\widehat{g^{-1}}_{\sigma \sigma ^{\prime }}.
\end{eqnarray}

$A_{k=0}$ is fixed in the low-energy limit but the other constants $%
a_{\uparrow }$,$a_{\downarrow },b$ are non-universal and depend on
high-energy processes. When there is spin-charge separation, the exponent
for $4k_{F}$ oscillations and the constant $A_{k=0}$ are related by the
equation $2+\alpha (4k_{F})=4A_{k=0}$; { in the general case,} we find: 
\[
2+\alpha (2k_{F\uparrow }+2k_{F\downarrow })=4A_{k=0} 
\]

The derivation of the exponents is done in exactly the same manner as in the
spin-charge separated LL.

Spin-spin correlation functions are: 
\begin{eqnarray}
&<&S_{z}(0)S_{z}(x)>=\frac{\sum_{\sigma \sigma ^{\prime }}\left( \sigma
\sigma ^{\prime }\widehat{g^{-1}}_{\sigma \sigma ^{\prime }}\right) }{2(\pi
x)^{2}}  \nonumber \\
&&+\sum_{\sigma }\frac{\cos 2k_{\sigma }x}{\left| x\right| ^{2\widehat{g}%
_{\sigma \sigma }^{-1}}},
\end{eqnarray}

\begin{eqnarray}
&<&S^{+}(0)S^{-}(x)>=\frac{\cos (k_{F\uparrow }+k_{F\downarrow })x}{\left|
x\right| ^{\gamma }}, \\
\gamma &=&\left[ g_{\uparrow \downarrow }^{-1}-g_{\uparrow \downarrow }+%
\frac{1}{2}\sum_{\sigma }(\widehat{g}^{-1}+\widehat{g})_{\sigma \sigma
}\right] .
\end{eqnarray}

Electronic Green functions decay as: 
\begin{equation}
<\Psi _{\sigma }(0)\Psi _{\sigma }^{+}(x)>=\frac{\exp ik_{\sigma }x}{\left|
x\right| ^{1+\alpha (\sigma )}};\;1+\alpha _{F}(\sigma )=\frac{1}{2}(%
\widehat{g}+\widehat{g}^{-1})_{\sigma \sigma }.
\end{equation}
For bosons the exponent is modified as: $1+\alpha _{B}(\sigma )=\frac{1}{2}%
\widehat{g}_{\sigma \sigma }.$ These exponents are derived with the
bosonization formulas but can also be found by plasma analogy.

\subsubsection{The charge matrix: a summary.}

For the two-component LL there are three interesting situations which we
summarize below:

i) in the general case, the (symmetric) $\widehat{g}$ matrix has arbitrary
entries; there is no spin-charge separation but a more general two-mode
separation: 
\begin{equation}
\widehat{g}=\left( 
\begin{array}{ll}
\lambda & \mu \\ 
\mu & \lambda ^{\prime }
\end{array}
\right) .
\end{equation}
As will be shown below, the Hubbard model in a magnetic field can be
described by such a theory.

ii) the $\widehat{g}$ matrix has a $Z_{2}$ symmetry;
 {this case pertains to} 
spin-charge separation: 
\begin{equation}
\widehat{g}=\left( 
\begin{array}{ll}
\lambda & \mu \\ 
\mu & \lambda
\end{array}
\right) ,\;K_{\rho }=\frac{1}{\lambda +\mu },\;K_{\sigma }=\frac{1}{\lambda
-\mu }
\end{equation}
Indeed the symmetry under the exchange of up and down spins implies that the
normal modes of the charge matrix are just the charge and spin modes. The LL
parameters are then the eigenvalues of the inverse of the charge matrix.
This situation describes models with spin-charge separation but with a spin
anisotropy, for instance a Hubbard model to which one would add some Ising
term $\sum_{n}S_{z}(n)S_{z}(n+1)$.

iii) the $\widehat{g}$ matrix corresponds to a $SU(2)$ symmetric case
 ($K_{\sigma }=1$): 
\begin{equation}
\widehat{g}=\left( 
\begin{array}{ll}
\mu +1 & \mu \\ 
\mu & \mu +1
\end{array}
\right) ,\qquad K_{\rho }=\frac{1}{2\mu +1},K_{\sigma }=1
\end{equation}
This situation describes the low-energy limit of the Hubbard model. It is
noteworthy that the wavefunctions $\Psi \left[ \widehat{g}\right] $ for that
sub-case were used in a variational approach of the 1D $t-J$ model giving
very good results although it was not realized they were the exact ground
states of the gaussian model\cite{hel}. {The reason why it is so is now
transparent.}
\subsection{Elementary excitations for the generalized LL.}

\label{ch-33}

We now consider the excitations of the general bosonic and fermionic LL with
or without spin-charge separation. Let us inject particles in the Luttinger
liquid. In real space this is described by the operator: 
\begin{equation}
V(x)=\exp -i\sqrt{\pi }\sum_{\sigma }\left( Q_{\sigma }\Theta _{\sigma
}(x)-J_{\sigma }\Phi _{\sigma }(x)\right) .
\end{equation}
Fractionalization stems from two decouplings: chiral separation and a
separation for the internal quantum number generalizing spin-charge
separation. In terms of the normal modes fields $\Theta _{\tau }$ and $\Phi
_{\tau }$ $(\tau =1,2$): 
\begin{equation}
V=\exp -i\sqrt{\pi }\sum_{\tau }\left( (\sum_{\sigma }P_{\sigma \tau
}Q_{\sigma })\Theta _{\tau }-(\sum_{\sigma }P_{\sigma \tau }J_{\sigma })\Phi
_{\tau }\right) .
\end{equation}
The chiral fields are: 
\begin{equation}
\Theta _{\tau ,\pm }(x)=\Theta _{\tau }(x)\mp \Phi _{\tau }(x)/K_{\tau },
\end{equation}
and therefore: 
\begin{equation}
V(x)=\prod_{\tau ,\pm }\exp -i\sqrt{\pi }Q_{\tau ,\pm }\Theta _{\tau ,\pm
}(x).
\end{equation}
This expression explicitly shows a decoupling into four components. We have
defined in the above the chiral charges: 
\begin{equation}
Q_{\tau ,\pm }=\frac{1}{2}\left[ (\sum_{\sigma }P_{\sigma \tau }Q_{\sigma
})\pm K_{\tau }(\sum_{\sigma }P_{\sigma \tau }J_{\sigma })\right] .
\end{equation}
The following operators are exact eigenstates of each chiral
hamiltonian $H_{\pm ,\tau }$ $\tau =1,2$:

\begin{eqnarray}
V_{\tau }^{\pm }(Q_{\tau ,\pm },q) &=&\int dx\exp iqx\exp -i\sqrt{\pi }%
Q_{\tau ,\pm }\Theta _{\tau ,\pm }(x), \\
q &=&\frac{2\pi n}{L}\mp \frac{2\pi }{L}\frac{Q_{\tau ,\pm }^{2}}{K_{\tau }}.
\end{eqnarray}
The chiral charges correspond to the charge and spin carried by each of
these excitations up to a normalization factor: 
\begin{equation}
\left[ \widehat{Q}_{\sigma },V_{\tau }^{\pm }(Q_{\tau ,\pm },q)\right]
=Q_{\tau ,\pm }P_{\sigma \tau }V_{\tau }^{\pm }(Q_{\tau ,\pm },q),
\end{equation}
which implies that the charge and spin of $V_{\tau }^{\pm }(Q_{\tau ,\pm
},q) $ are:

\begin{eqnarray}
Q &=&Q_{\tau ,\pm }\left( \sum_{\sigma }P_{\sigma \tau }\right) ,  \label{c1}
\\
S_{z} &=&Q_{\tau ,\pm }\left( \frac{1}{2}\sum_{\sigma }\sigma P_{\sigma \tau
}\right) .  \label{c2}
\end{eqnarray}
Thus for an arbitrary charge matrix, fractional excitations carry
both charge and spin. However the ratio of charge to spin is constant for
each given mode $\tau =1,2$: $\left( \frac{1}{2}\sum_{\sigma }\sigma
P_{\sigma \tau }\right) Q=S_{z}\left( \sum_{\sigma }P_{\sigma \tau }\right) $%
; of course the phonons associated to each mode mix charge and spin in
exactly the same proportions since:

\begin{eqnarray}
\rho _{\tau }(x) &=&\left( \frac{1}{2}\sum_{\sigma }P_{\sigma \tau }\right)
\rho _{c}(x)+\left( \frac{1}{2}\sum_{\sigma }\sigma P_{\sigma \tau }\right)
\rho _{s}(x)  \nonumber \\
&=&\left( \frac{1}{2}\sum_{\sigma }P_{\sigma \tau }\right) \rho
_{c}(x)+\left( \sum_{\sigma }P_{\sigma \tau }\right) s_{z}(x)
\end{eqnarray}
where $s_{z}(x)$ is a spin density.

It is convenient to define the charge to spin ratio: 
\begin{equation}
r=\frac{Q}{2S_{z}},
\end{equation}
for each mode. Unitary implies that if for the first mode: 
\begin{equation}
r=\frac{Q}{2S_{z}}=p
\end{equation}
then for the second mode: 
\begin{equation}
r=\frac{Q}{2S_{z}}=-\frac{1}{p}.
\end{equation}
where $p$ is arbitrary. Note that for
a Fermi liquid these ratios are $r=\pm 1$ (we are characterizing Landau
quasiparticles (or holes) of either spin) and when spin-charge separation is
realized the ratio is either $r=0$ or $r=\pm \infty $. 

Let us give the elementary excitations. The simplest case is
that of bosons: 
\[
Q_{\tau ,\pm }=Q_{\uparrow }\left( \frac{P_{\uparrow \tau }}{2}\right)
+Q_{\downarrow }\left( \frac{P_{\downarrow \tau }}{2}\right) 
\]
\begin{equation}
+\frac{J_{\uparrow }}{2}\left( \pm P_{\uparrow \tau }K_{\tau }\right) +\frac{%
J_{\downarrow }}{2}\left( \pm P_{\downarrow \tau }K_{\tau }\right) .
\end{equation}

To simplify the notation, we have only written a single line, but $Q_{\tau ,\pm
} $ and the other entries should be read as four-vectors. $Q_{\sigma }$ and $%
\frac{J_{\sigma }}{2}$ are arbitrary independent integers, which shows that
the states $V_{\tau }^{\pm }(\widetilde{Q}_{\tau ,\pm },q)$ where $%
\widetilde{Q}_{\tau ,\pm }=\frac{P_{\uparrow \tau }}{2},\frac{P_{\downarrow
\tau }}{2},\pm P_{\uparrow \tau }K_{\tau }$ or $\pm P_{\downarrow \tau
}K_{\tau }$ are elementary excitations. As an illustration let us consider
the simple case of a $Z_{2}$ symmetric charge matrix for bosons which have a
pseudo-spin index. The unitary matrix $P$ is: 
\begin{equation}
P_{\sigma \tau }=\left( 
\begin{tabular}{ll}
$1/\sqrt{2}$ & $1/\sqrt{2}$ \\ 
$1/\sqrt{2}$ & $-1/\sqrt{2}$%
\end{tabular}
\right)
\end{equation}
Then it follows from eq.(\ref{c1}) and (\ref{c2}) that for mode $\tau =1$
(the charge mode) $Q=\sqrt{2}Q_{\tau ,\pm }$ and $S_{z}=0$; for mode $\tau
=2 $ $Q=0$ but $S_{z}=Q_{\tau ,\pm }/\sqrt{2}$. If we take 
these normalizations into account: 
\begin{eqnarray}
\left( 
\begin{tabular}{l}
$Q_{+}$ \\ 
$Q_{-}$ \\ 
$S_{+}$ \\ 
$S_{-}$%
\end{tabular}
\right) &=&Q_{\uparrow }\left( 
\begin{tabular}{l}
$1/2$ \\ 
$1/2$ \\ 
$1/4$ \\ 
$1/4$%
\end{tabular}
\right) +Q_{\downarrow }\left( 
\begin{tabular}{l}
$1/2$ \\ 
$1/2$ \\ 
$-1/4$ \\ 
$-1/4$%
\end{tabular}
\right)  \nonumber \\
&&+\frac{J_{\uparrow }}{2}\left( 
\begin{tabular}{l}
$K_{c}$ \\ 
$-K_{c}$ \\ 
$K_{s}/2$ \\ 
$-K_{s}/2$%
\end{tabular}
\right) +\frac{J_{\downarrow }}{2}\left( 
\begin{tabular}{l}
$K_{c}$ \\ 
$-K_{c}$ \\ 
$-K_{s}/2$ \\ 
$K_{s}/2$%
\end{tabular}
\right)
\end{eqnarray}

Once again we find a charge $1/2$ particle and a charge $K_{c}$ Laughlin
quasiparticle as for the bosonic spinless LL. But in addition we find new
states resulting from a fractionalization of ''pseudo-spin'' for the bosons
(half-spinons for instance).

For the fermionic LL the elementary excitations are obtained by the
equation: 
\begin{eqnarray}
Q_{\tau ,\pm } &=&Q_{\uparrow }\left( P_{\uparrow \tau }\frac{1\pm K_{\tau }%
}{2}\right) +Q_{\downarrow }\left( P_{\downarrow \tau }\frac{1\pm K_{\tau }}{%
2}\right)  \nonumber \\
&&+n_{\uparrow }\left( \mp P_{\uparrow \tau }K_{\tau }\right) +n_{\downarrow
}\left( \mp P_{\downarrow \tau }K_{\tau }\right) ,
\end{eqnarray}
where we have resolved the constraint: $Q_{\sigma }-J_{\sigma }=2n_{\sigma }$%
. {This fully characterizes the low-energy elementary excitations 
of a LL in a magnetic field (see below)}.


\subsection{Application to the Hubbard model.}

\label{ch-34}

To illustrate the previous results we discuss the Hubbard model in one dimension. The model was solved exactly by Bethe Ansatz by Lieb and Wu.
{In zero magnetic field, for repulsive 
($U>0$) interactions, a LL metallic phase
exists both for weak and strong coupling, except at half-filling.}
  For very large $U$ the spin-charge 
decoupling is valid at all energy scales. This
was shown by Ogata and Shiba who also found that the Bethe Ansatz ground
state then took a remarkable factorized form\cite{ogata}: it is the product
of a charge part (a Slater determinant for free fermions involving all
electrons ) and a Bethe wavefunction similar to that of the Heisenberg model
on a reduced lattice from which one has removed the holes 
\begin{eqnarray}
&&\Psi _{Hubbard}(x_{i},\sigma _{i})  \nonumber \\
&=&\det (\exp ik_{j}r_{i},\left| k_{j}\right| \leq k_{F})\;\Psi
_{Heisenberg}(y_{i},\sigma _{i})
\end{eqnarray}
($y_{i}$ is the coordinate in the reduced lattice of particle $i$ whose real
position is $x_{i}$).

It is instructive to compare it to the two-component Jastrow wavefunctions
which are also explicitly spin-charge decoupled. The Slater determinant is
rewritten as (in terms of the circular coordinates $z$): 
\begin{equation}
\Psi _{Hubbard}(x_{i},\sigma _{i})=\prod_{i<j}\left( z_{i}-z_{j}\right)
\;\Psi _{Heisenberg}(y_{i},\sigma _{i})
\end{equation}
This is to be compared with: 
\begin{eqnarray}
\Psi &=&\prod_{i<j}\left| z_{i}-z_{j}\right| ^{1/2K_{c}}\;\prod_{i<j}\left|
z_{i}-z_{j}\right| ^{\sigma _{i}\sigma _{j}/2K_{s}}\;  \nonumber \\
&&\prod_{i<j}\left\{ \left( \frac{(z_{i}-z_{j})}{\left| z_{i}-z_{j}\right| }%
\right) ^{\delta _{\sigma _{i}\sigma _{j}}}\exp i\frac{\pi }{2}sgn(\sigma
_{i}-\sigma _{j})\right\}
\end{eqnarray}
Or if we separate spin and charge:

\begin{eqnarray}
\Psi _{c} &=&\prod_{i<j}\left[ \left| z_{i}-z_{j}\right| ^{1/2K_{c}}\left( 
\frac{(z_{i}-z_{j})}{\left| z_{i}-z_{j}\right| }\right) ^{1/2}\right] , \\
\Psi _{s} &=&\prod_{i<j}[\left| z_{i}-z_{j}\right| ^{\sigma _{i}\sigma
_{j}/2K_{s}}  \nonumber \\
&&\times \left( \frac{(z_{i}-z_{j})}{\left| z_{i}-z_{j}\right| }\right)
^{\sigma _{i}\sigma _{j}/2}\exp i\frac{\pi }{2}sgn(\sigma _{i}-\sigma _{j})].
\end{eqnarray}
The spin part of the Laughlin ground state is just the Haldane-Shastry
wavefunction if $K_{s}=1$ (rotational invariance) which has the same
large-distance physics as the Heisenberg ground state. We can also determine 
$K_{c}$ without any computation by just reading off its value from the
wavefunctions: the charge parts of the two wavefunctions coincide if $%
K_{c}=1/2$ which indeed is the known value of the LL parameter for large $U$.

Bethe Ansatz gives the spectrum and the eigenstates; however
it is very difficult to compute correlation functions. An important advance
came however with the works of Frahm and Korepin who used CFT in conjunction
with Bethe Ansatz to compute critical exponents\cite{frahm}. If a theory is
conformally invariant, one can show that the finite-size energies of
excitations are directly related to their operator dimension (which is
one-half of the anomalous dimension of their correlation function). By using
Woynarovich's Bethe Ansatz calculations for the finite-size spectrum to 
order $1/L$
which he computed within a so-called ''dressed charge matrix formalism''\cite
{woy}, Frahm and Korepin were able to extract critical exponents for the
correlation functions of the Hubbard model. In particular they found that in
the presence of a magnetic field, spin-charge separation was not realized.
Penc and Solyom later showed that in $1/L$ the same spectrum derived by
Woynarovich could be expressed in terms of a generalized Tomonaga-Luttinger
model with interactions described in the g-ology framework; using equation
of motion methods they also derived the anomalous exponents\cite{penc}.
 These two
approaches {give little insight into the nature} of the elementary
 excitations:
how are the holon and spinon modified as a function of microscopic
parameters? The description of spin-charge separation (or its absence) is
not transparent either: the dressed charge matrix tell us little about 
spin-charge separation; its changes are not easy to relate to that 
property. This is to be contrasted with our charge
matrix formalism in which spin-charge separation is directly connected to a
symmetry of the charge matrix $\widehat{g}$ ($Z_{2}$ symmetry). 
We will show that the ''dressed charge matrix'' of Bethe Ansatz
and the charge matrix $\widehat{g}$ are in fact related: the inverse of the
symmetric charge matrix is roughly the square of the $Z$ matrix. We will
proceed in the following manner: we will show that Woynarovich's finite-size
spectrum is identical to that of our generalized LL. {This yields
the charge matrix $\widehat{g}$ in terms of the dressed charge
matrix $Z$ and gives us both the anomalous exponents
and the fractional excitations}
 since we already derived them for the generalized
LL. The relation of our charge matrix formalism to Penc and Solyom g-ology
approach is the following: it can be understood as a bosonization of their
generalized Tomonaga-Luttinger model; it is much simpler however to work
directly within the gaussian hamiltonian framework. Our approach has several
advantages in addition to making an explicit contact with the  seemingly
unrelated physics of Laughlin states: (a) we avoid an ambiguity in the
determination of anomalous exponents in Frahm's and Korepin approach;\cite
{f16} (b), we can give the nature of elementary excitations ( and show
that they are fractional states in the first place ) and (c) we are able
to give a clear criterion of spin-charge separation.

Woynarovich's finite-size spectrum in Frahm and Korepin's notations is the
following: 
\begin{eqnarray}
&&E({ \Delta N},{ D,}N_{c}^{\pm },N_{s}^{\pm })-E_{0}  \nonumber \\
&=&\frac{2\pi }{L}\left[
v_{c}(N_{c}^{+}+N_{c}^{-})+v_{s}(N_{s}^{+}+N_{s}^{-})\right]  \nonumber \\
&&+\frac{2\pi }{L}\left[ \frac{1}{4}{ \Delta N}^{T}(Z^{-1})^{T}{ V}%
Z^{-1}{ \Delta N+D}^{T}Z{ V}Z^{T}{ D}\right]  \nonumber \\
&&+O(\frac{1}{L}),
\end{eqnarray}

\begin{eqnarray}
&&P({ \Delta N},{ D,}N_{c}^{\pm },N_{s}^{\pm })-P_{0}  \nonumber \\
&=&\frac{2\pi }{L}\left[
v_{c}(N_{c}^{+}-N_{c}^{-})+v_{s}(N_{s}^{+}-N_{s}^{-})\right]  \nonumber \\
&&+\frac{2\pi }{L}\left[ { \Delta N}^{T}{ D}\right]
+2D_{c}k_{F\uparrow }+2(D_{c}+D_{s})k_{F\downarrow }.
\end{eqnarray}
$k_{F\uparrow }=\frac{2\pi }{L}N_{\uparrow }$ and $k_{F\downarrow }=\frac{%
2\pi }{L}N_{\downarrow }$ are the Fermi momentum for particles of spin up
and spin down. The energy and the momentum are those of a state with the
(integer) quantum numbers (${ \Delta N},{ D,}N_{c}^{\pm },N_{s}^{\pm
} $); there are two modes indexed by $c$ and $s$: these two modes {\it do not%
} in general correspond to charge and spin. $Z$ is a $2$ by $2$
matrix: 
\begin{equation}
Z=\left( 
\begin{tabular}{ll}
$Z_{cc}$ & $Z_{cs}$ \\ 
$Z_{sc}$ & $Z_{ss}$%
\end{tabular}
\right)
\end{equation}
and ${ \Delta N}$ and ${ D}$ are two-vectors: ${ \Delta N=(}%
N_{c}=N_{\uparrow }+N_{\downarrow },N_{s}=N_{\downarrow })$ and ${ D=(}%
D_{c},D_{s})$.\cite{f17} In these expressions $N_{c/s}^{+-}$ are integers:
they are simply the modulus of phonon momenta in units of $2\pi /L$ for the
two modes $c$ and $s$; the index $\pm $ refers to the sign of the momentum.
The phonon velocities for the two modes are $(v_{c},v_{s})$.

The spectrum of the general gaussian model $H\left[ u_{\tau },\widehat{g}%
\right] $ is: 
\begin{eqnarray}
&&E(Q_{\sigma },J_{\sigma },N_{\tau }^{\pm })  \nonumber \\
&=&\frac{2\pi }{L}\left[ v_{\tau =1}(N_{\tau =1}^{+}+N_{\tau
=1}^{-})+v_{\tau =2}(N_{\tau =2}^{+}+N_{\tau =2}^{-})\right]  \nonumber \\
&&+\frac{\pi }{2L}\sum_{\tau =1,2}v_{\tau }\left( \frac{Q_{\tau }^{2}}{%
K_{\tau }}+K_{\tau }J_{\tau }^{2}\right) , \\
&&P(Q_{\sigma },J_{\sigma },N_{\tau }^{\pm })  \nonumber \\
&=&\frac{2\pi }{L}\left[ (N_{\tau =1}^{+}-N_{\tau =1}^{-})+(N_{\tau
=2}^{+}-N_{\tau =2}^{-})\right]  \nonumber \\
&&+\sum_{\sigma =\uparrow \downarrow }\frac{\pi Q_{\sigma }}{L}J_{\sigma
}+k_{F\sigma }J_{\sigma }.
\end{eqnarray}
$N_{\tau =1}^{\pm },N_{\tau =2}^{\pm }$ are again the moduli of phonon
momenta. The charges and currents $(Q_{\tau },J_{\tau })$ are related to $%
(Q_{\sigma },J_{\sigma })$ by $Q_{\tau }=P_{\sigma \tau }Q_{\sigma }$ and $%
J_{\tau }=P_{\sigma \tau }J_{\sigma }$. We can now identify the parameters
of both theories:

\begin{eqnarray}
J_{\uparrow } &=&2D_{c},\;J_{\downarrow }=2(D_{c}+D_{s}),\; \\
Q_{\uparrow } &=&N_{\uparrow },\;Q_{\downarrow }=N_{\downarrow },\;N_{\tau
}^{\pm }=N_{c/s}^{\pm }, \\
v_{\tau } &=&v_{c/s}.
\end{eqnarray}
The zero modes can be identified term by term; it is sufficient to consider
the current terms to uniquely determine the charge matrix. The charge
zero modes yield extra relations which lead to the very
same expression for $\widehat{g}$. Indeed expanding the squares
gives:

\begin{eqnarray}
K_{1}P_{\uparrow 1}^{2} &=&(Z_{cc}-Z_{sc})^{2},  \nonumber \\
K_{1}P_{\downarrow 1}^{2} &=&(Z_{sc})^{2},  \nonumber \\
K_{1}P_{\uparrow 1}P_{\downarrow 1} &=&(Z_{cc}-Z_{sc})Z_{sc},  \nonumber \\
K_{2}P_{\uparrow 2}^{2} &=&(Z_{cs}-Z_{ss})^{2},  \nonumber \\
K_{2}P_{\downarrow 2}^{2} &=&(Z_{ss})^{2},  \nonumber \\
K_{2}P_{\uparrow 2}P_{\downarrow 2} &=&(Z_{cs}-Z_{ss})Z_{ss}.
\end{eqnarray}

Since: 
\begin{equation}
\widehat{g^{-1}}_{\sigma \sigma ^{\prime }}=\sum_{\tau }K_{\tau }P_{\sigma
\tau }P_{\sigma ^{\prime }\tau }
\end{equation}
it follows that the inverse of the charge matrix is: 
\begin{eqnarray}
\widehat{g^{-1}}_{\uparrow \uparrow }
&=&(Z_{cc}-Z_{sc})^{2}+(Z_{cs}-Z_{ss})^{2}  \nonumber \\
\widehat{g^{-1}}_{\downarrow \downarrow } &=&(Z_{sc})^{2}+(Z_{ss})^{2} 
\nonumber \\
\widehat{g^{-1}}_{\uparrow \downarrow } &=&\widehat{g^{-1}}_{\downarrow
\uparrow }=(Z_{cc}-Z_{sc})Z_{sc}+(Z_{cs}-Z_{ss})Z_{ss}
\end{eqnarray}
We define the matrix $\widetilde{Z}$ obtained from the dressed charge matrix 
$Z$ by subtracting the second line from the first: 
\begin{equation}
\widetilde{Z}=\left( 
\begin{tabular}{ll}
$Z_{cc}-Z_{sc}$ & $Z_{cs}-Z_{ss}$ \\ 
$Z_{sc}$ & $Z_{ss}$%
\end{tabular}
\right)
\end{equation}
Then: 
\begin{equation}
\widehat{g}^{-1}=\widetilde{Z}\widetilde{Z}^{T}
\end{equation}

{This is the most important result of the present section: 
the low-energy properties of the Hubbard model are expressed in terms of
quantities which can be computed from the microscopic
parameters and  spin-charge separation 
simply follows from the 
$Z_{2}$ symmetry of our charge matrix. In the framework of 
the dressed charge matrix $Z$ approach, the second feature is not 
easily decoded from the structure of $Z$
which is then triangular} 
with some relations between its matrix
elements whose physical interpretation is quite unclear.\cite{f18} We can 
use the results of the previous sections on the elementary excitations
and those on the various properties of the charge matrix hamiltonian such as
 the Drude
peak, the susceptibility, the anomalous exponents. In particular the new modes
replacing the spin and charge modes are simply the eigenvectors of the
charge matrix.

The charge matrix is obtained by inversion: 
\[
\widehat{g}=\frac{1}{\left( \det Z\right) ^{2}} 
\]
\begin{equation}
\times \left( 
\begin{tabular}{ll}
$\widehat{g^{-1}}_{\downarrow \downarrow }$ & $-\widehat{g^{-1}}_{\uparrow
\downarrow }$ \\ 
$-\widehat{g^{-1}}_{\uparrow \downarrow }$ & $\widehat{g^{-1}}_{\uparrow
\uparrow }$%
\end{tabular}
\right) .  \label{g}
\end{equation}

Term by term identification of the charge zero modes $Q_{\sigma }$ would
lead to exactly the same expression for $\widehat{g}$; indeed:

\begin{eqnarray}
\frac{1}{K_{1}}P_{\uparrow 1}^{2} &=&\frac{(Z_{ss})^{2}}{\left( \det
Z\right) ^{2}},  \nonumber \\
\frac{1}{K_{1}}P_{\downarrow 1}^{2} &=&\frac{(Z_{cs}-Z_{ss})^{2}}{\left(
\det Z\right) ^{2}},  \nonumber \\
\frac{1}{K_{1}}P_{\uparrow 1}P_{\downarrow 1} &=&\frac{-(Z_{cs}-Z_{ss})Z_{ss}%
}{\left( \det Z\right) ^{2}},  \nonumber \\
\frac{1}{K_{2}}P_{\uparrow 2}^{2} &=&\frac{(Z_{sc})^{2}}{\left( \det
Z\right) ^{2}}^{2},  \nonumber \\
\frac{1}{K_{2}}P_{\downarrow 2}^{2} &=&\frac{(Z_{cc}-Z_{sc})^{2}}{\left(
\det Z\right) ^{2}},  \nonumber \\
\frac{1}{K_{2}}P_{\uparrow 2}P_{\downarrow 2} &=&\frac{-(Z_{cc}-Z_{sc})Z_{sc}%
}{\left( \det Z\right) ^{2}},
\end{eqnarray}
and since $\widehat{g}_{\sigma \sigma ^{\prime }}=\sum_{\tau }K_{\tau
}^{-1}P_{\sigma \tau }P_{\sigma ^{\prime }\tau }$ one recovers equ.(\ref{g}%
). As it should be, one can check that the anomalous exponents predicted for 
$H\left[ u_{\tau },\widehat{g}\right] $ agree then completely with Frahm and
Korepin's results.

{Let us illustrate these results in two situations,  one with 
spin-charge separation, the other without. From these, we can exhibit
the criterion for spin-charge separation within the 
matrix formalism.}

In the presence of spin-charge separation, Frahm and Korepin find that the
dressed charge matrix $Z$ is $Z=\left( 
\begin{tabular}{ll}
$Z_{cc}=\xi $ & $Z_{cs}=0$ \\ 
$Z_{sc}=\xi /2$ & $Z_{ss}=1/\sqrt{2}$%
\end{tabular}
\right) $. This implies that $\widehat{g}$ and its inverse $\widehat{g}%
^{-1}$ are:

\begin{eqnarray}
\widehat{g}^{-1} &=&\left( 
\begin{tabular}{ll}
$\frac{\xi ^{2}}{4}+\frac{1}{2}$ & $\frac{\xi ^{2}}{4}-\frac{1}{2}$ \\ 
$\frac{\xi ^{2}}{4}-\frac{1}{2}$ & $\frac{\xi ^{2}}{4}+\frac{1}{2}$%
\end{tabular}
\right) \\
\widehat{g} &=&\left( 
\begin{tabular}{ll}
$\frac{1}{\xi ^{2}}+\frac{1}{2}$ & $\frac{1}{\xi ^{2}}-\frac{1}{2}$ \\ 
$\frac{1}{\xi ^{2}}-\frac{1}{2}$ & $\frac{1}{\xi ^{2}}+\frac{1}{2}$%
\end{tabular}
\right)
\end{eqnarray}
The charge matrix explicitly exhibits spin-charge separation and takes the
form characteristic of $SU(2)$ symmetry. The eigenvalues of $\widehat{g}%
^{-1} $ are $K_{c}=\frac{\xi ^{2}}{2}$ and $K_{s}=1$.

The $Z$ matrix can also be explicitly computed 
in the limit of infinite repulsion with a magnetic field close
to the critical field $h_{c}$ for which all the spins are polarized (i.e.
close to the ferromagnetic phase). In terms of the parameter 
\begin{equation}
\delta =\sqrt{\frac{h_{c}-h}{h_{c}}}
\end{equation}
the dressed charge matrix is: 
\begin{equation}
Z=\left( 
\begin{tabular}{ll}
$1$ & $0$ \\ 
$\frac{2}{\pi }\delta $ & $1-\frac{1}{\pi }\delta $%
\end{tabular}
\right)
\end{equation}
which implies that the (inverse of the) charge matrix $\widehat{g}$ is: 
\[
\widehat{g}^{-1} = 
\]
\begin{equation}
\left( 
\begin{tabular}{ll}
$\left( 1-\frac{1}{\pi }\delta \right) ^{2}+\left( 1-\frac{2}{\pi }\delta
\right) ^{2}$ & $\left( 1-\frac{2}{\pi }\delta \right) \left( \frac{2}{\pi }%
\delta \right) -\left( 1-\frac{1}{\pi }\delta \right) ^{2}$ \\ 
$\left( 1-\frac{2}{\pi }\delta \right) \left( \frac{2}{\pi }\delta \right)
-\left( 1-\frac{1}{\pi }\delta \right) ^{2}$ & $\left( 1-\frac{1}{\pi }%
\delta \right) ^{2}+\left( \frac{2}{\pi }\delta \right) ^{2}$%
\end{tabular}
\right) .
\end{equation}
This expression shows explicitly the breakdown of  spin-charge separation 
(except for $\delta =\pi /4$).


\section{Conclusions and Perspectives.}

\label{ch-4}

The main goal of our paper was to establish and describe fractional
excitations for the Luttinger liquid within the bosonization scheme. {\ bf Bethe
Ansatz gives exact eigenstates 
and shows the existence
of some fractional excitations: however their
description is quite complex in that framework and it is unclear how to 
generate systematically a complete set of excitations.} 
{In the low-energy limit, the 
Luttinger liquid approach allows a very precise
characterization of the fractional states already known from exact
solutions but what's more allows us to discover novel fractional}
excitations which may carry irrational charges (the 1D Laughlin
quasiparticle, the hybrid state). In section \ref{ch-two}
the low-energy spectrum of Luttinger liquids can be reinterpreted
in terms of fractional states: for instance, the particle-hole continuum 
consists of a Laughlin quasiparticles-quasihole continuum. The quasiparticle
perspective clarifies many properties of
the LL: the renormalization of the current operator is a
direct consequence of fractionalization; for spin chains, we present the correct
description of the spinon excitation in the generic case of a
violation of $SU(2)$ invariance. We also  show that the $S_{z}=0$
continuum of spin chains involves the analogs of Laughlin
quasiparticles. In section \ref{ch-3} we describe fractional
excitations such as the holon or the spinon for the Luttinger liquid with
spin; we also present a generalization of the gaussian theory valid for
Luttinger liquids without spin-charge separation and display
in that situation the new fractional states replacing the holon and spinon
(see \ref{ch-33}).

An important test, of course, would be to observe experimentally (or
numerically) all these fractional states. Although the existence of the
holon and the spinon was ascertained theoretically quite a long time ago\cite
{lieb} no experiment  has yet allowed
their detection: in fact, the property of spin-charge separation itself is
not yet established experimentally. The observation of two of
the fractional states discussed in this paper would be particularly
important: the LL Laughlin quasiparticle and the hybrid state. Indeed they
may assume irrational charges. The precise spectroscopy of fractional
excitations we have done in this paper allows us to determine which
processes are involved in their creation: for the Laughlin states, current
probes are needed, while the hybrid particle is created by addition of an
odd number of electrons. For Laughlin quasiparticles shot noise is likely
an adequate probe: the shot noise coefficient for Luttinger
liquids can be computed exactly and is predicted to be equal to $K$;\cite
{ludwig} in the two-dimensional electron gas at filling $\nu =1/3$ this
yields a charge $1/3$\cite{glattli}. The latter situation involves Wen's
chiral Luttinger liquid. The identification of the shot noise coefficient
with the charge of a carrier has been debated because the coefficient one
measures might actually be the conductance rather than a quasiparticle
charge (at $\nu =1/3$ the conductance also assumes the value $1/3$).
For the non-chiral Luttinger liquid our spectroscopy of fractional states
allows to resolve that ambiguity: the shot noise coefficient is indeed
identical to the conductance $K$ of the LL but the (backscattering)
current-current correlation function measured in shot noise involves
charge $K$ excitations, because charge $K$ LL Laughlin quasiparticles are
precisely generated by current excitations. These
might thus be detected in any physical realization of the Luttinger
liquid: quantum wires, or possibly nanotubes. An intriguing possibility is
also suggested by recent experiments of tunneling at the edge of a
two-dimensional gas in a magnetic field\cite{grayson}. $I-V$ characteristics
measured at the edge showed very surprising non-Fermi liquid behaviour
compatible with a chiral LL with unquantized LL parameter $K$; the
$I-V$ curves evolve smoothly when one varies the filling fraction and
do not show a plateau structure. There seems to be a continuum of
Luttinger liquids living at the edge: this caused quite a stir
because the chiral LL theory can presumably be derived only for
incompressible filling fractions. These puzzling results are sofar
unexplained, but if an interpretation in terms of a single-boson mode chiral
LL with unquantized parameter $K$ can be in some manner justified, according
to the results given in the present paper this would imply that there exists
charge $K$ Laughlin quasiparticles in that experimental setting: such
a chiral LL is identical to the chiral half of the non-chiral gaussian
hamiltonian considered throughout our paper.

The authors wish to acknowledge the late Heinz Schulz for insightful remarks
 on the results of this
paper. We
also thank Vincent Pasquier for a discussion on the
Calogero-Sutherland model. Bernard Jancovici gave us information on one
dimensional plasmas for which we are grateful.

\appendix 

\section{Dispersion of the fractional states.}

We show here that $V_{Q_{\pm }}^{\pm }(q)|\Psi _{0}>$ (where $|\Psi
_{0}>$ is the interacting ground state) is an exact eigenstate of the chiral
hamiltonian $H_{\pm }$ with energy: 
\begin{equation}
E(Q_{\pm },\overline{q_{n}})=\left[ u\left| \overline{q_{n}}\right| +\frac{%
\pi u}{2L}\frac{Q_{\pm }^{2}}{K}\right] .
\end{equation}

Let us rewrite the state considered $V_{Q_{\pm }}^{\pm }(q_{n})|\Psi _{0}>$
in terms of zero modes and phonon operators (we use equ. (\ref{theta}), (\ref
{phi}), (\ref{p1}) and (\ref{p2})): 
\[
V_{Q_{\pm }}^{\pm }(x)|\Psi _{0}>=\exp iQ_{\pm }\left[ \mp \frac{2\pi }{K}%
\widehat{Q}_{\pm }x-\sqrt{\pi }\left( \Theta _{0}\mp \frac{\Phi _{0}}{K}%
\right) \right] 
\]
\begin{equation}
\times \exp -i\sqrt{\pi }Q_{\pm }\sum_{n\neq 0}\Theta _{\pm ,n}\exp i\frac{%
2\pi n}{L}x|\Psi _{0}>
\end{equation}

Taking into account the fact that the operators $b_{q}$ annihilate the
ground state, it follows that: 
\[
V_{Q_{+}}^{+}(x)|\Psi _{0}>=\exp iQ_{+}\left[ -\frac{2\pi }{K}\widehat{Q}%
_{+}x-\sqrt{\pi }\left( \Theta _{0}-\frac{\Phi _{0}}{K}\right) \right] 
\]
\begin{equation}
\times \exp -i\sqrt{\pi }Q_{+}\sum_{n>0}\sqrt{\frac{L}{K\pi \left| n\right| }%
}b_{n}^{+}\exp i\frac{2\pi n}{L}x|\Psi _{0}>,
\end{equation}
with a similar expression for $V_{Q_{-}}^{-}(x)|\Psi _{0}>$ (the sum is then
over negative momentum phonons). Going back to reciprocal space: 
\begin{eqnarray*}
V_{Q_{\pm }}^{\pm }(q_{n})|\Psi _{0} &>&=\frac{1}{\sqrt{L}}%
\int_{0}^{L}dx\exp -i\frac{2\pi }{L}nx \\
&&\times \exp iQ_{+}\left[ -\sqrt{\pi }\left( \Theta _{0}\mp \frac{\Phi _{0}%
}{K}\right) \right]
\end{eqnarray*}
\begin{equation}
\times \exp -i\sqrt{\pi }Q_{+}\sum_{\pm p>0}\sqrt{\frac{L}{K\pi \left|
p\right| }}b_{p}^{+}\exp i\frac{2\pi p}{L}x|\Psi _{0}>,
\end{equation}
which shows that $V_{Q_{\pm }}^{\pm }(q_{n})|\Psi _{0}>$ only spans chiral
phonons with momentum $\pm n>0$ ; in other words this  state is obtained by
the action of the zero mode of the chiral field plus the
creation of phonons with {\it momenta of the same sign. }When one expands
the phonon exponential, the integral over position will select configuration
of phonons with identical total momentum $\overline{q_{n}}=\frac{2\pi n}{L}$%
. All these configurations consist of phonons of identical
chirality and total momentum which means that they are
eigenstates with the same eigenvalue of the appropriate chiral hamiltonian ($%
H_{+}$ or $H_{-}$). This is enough to prove that $V_{Q_{\pm }}^{\pm
}(q_{n})|\Psi _{0}>$ is an exact eigenstate of $H_{\pm }$:

\begin{eqnarray}
H_{\pm }V_{Q_{\pm }}^{\pm }(q_{n})|\Psi _{0} &>&=\left[
\sum_{q>0}u|q|a_{q}^{+}a_{q}+\frac{\pi u}{2L}\frac{\widehat{Q}_{\pm }^{2}}{K}%
\right] \\
\times V_{Q_{\pm }}^{\pm }(q_{n})|\Psi _{0} &>& \\
&=&\left[ u\left| \frac{2\pi n}{L}\right| +\frac{\pi u}{2L}\frac{Q_{\pm }^{2}%
}{K}\right] V_{Q_{\pm }}^{\pm }(q_{n})|\Psi _{0}>,
\end{eqnarray}
where $\overline{q_{n}}=\frac{2\pi n}{L}$ is the momentum due to phonons.

\clearpage 
\begin{figure}[tbp]
\caption{ Spectrum of the gaussian model with a bosonic Fock space in the
charge $Q$ sector. The energy at zero momentum is $\Delta (Q)=\pi uQ^2/(2L)$
as a function of $Q$ the number of particles added to the system. The
spectrum for fermionic Fock spaces is identical if $Q$ is an even integer.
The continuum is enclosed within the straight lines which are supported by
the parabolic enveloppe $\pi u (k/k_F)^2/(2L)$.}
\label{fig1}
\end{figure}

\clearpage 
\begin{figure}[tbp]
\caption{ Spectrum of the gaussian model for fermionic Fock spaces in the
case of charge sectors for which $Q$ is an odd integer. Notice that the
energy has now local minima for momenta $k=\pm 1,\pm 3,\pm 5, ...$ (in units
of $k_F$) instead of $k=0,\pm 2,\pm 4$,...}
\label{fig2}
\end{figure}

\clearpage 
\begin{figure}[tbp]
\caption{ Spectrum of the $XXZ$ spin chain for a transverse exchange $J$ and
anisotropy $\Delta$; the continuum is enclosed within
the curves $E(k)=(\pi \alpha /2)\mid \sin (k) \mid$ and $E(k)= \pi \alpha
\sin (k/2)$. The parameter $\alpha$ is related to the anisotropy $ \Delta =
\cos \theta$ by $\alpha = \sin \theta / \theta $. The linearized spectrum
found by bosonization is also shown.} 
\label{fig3}
\end{figure}


\begin{references}
\bibitem{f1}  We call these excitations describing collective density
fluctuations interchangeably phonons, or plasmons.
\bibitem{anomaly}  $c$ is a real number known as the conformal anomaly.
\bibitem{phase}  We omitted a phase factor corresponding to a momentum $\pi $
boost. It can be ignored after a bipartite transformation.
\bibitem{basis}  The zero mode basis (phonons and $U_{\pm}$ operators) forms a
complete family of states: this was proved by Haldane in his computation of
the grand canonical partition function of the Tomonaga-Luttinger model\cite
{haldane79}. Therefore $V_{Q,J}$ operators span the Fock space since $%
V_{Q,J} $ operators are products of Haldane's $U_{\pm}$ operators and of an
exponential of phonons. Since the fractional states $V_{Q_{\pm }}^{\pm }$
come from the decay of the $V_{Q,J}$ operators, they, along with the
phonons span the Fock space.
\bibitem{f4}  This is just the quantization of the
superfluid phase: the boson operator is $\Psi _{B}\propto \exp i\sqrt{\pi }%
\Theta $ and therefore $\sqrt{\pi }\left( \Theta (L)-\Theta (0)\right) $
must be an integer multiple of $\pi $. The total current around the ring is
defined as the integral $\int_{0}^{L}dx\;j_{R}(x)=\widehat{J}_{R}$;
therefore $\widehat{J}_{R}=\frac{uK}{\sqrt{\pi }}\left( \Theta (L)-\Theta
(0)\right) $ and we may then write $\widehat{J}_{R}=uKJ$ where $J$ is an
integer.
\bibitem{f6}  The previous equation assumes that both $Q$ and $n$ are
positive; if that is not so, it suffices to take their absolute value and
replace the operators by their conjugate: for instance $Q\longrightarrow
\left| Q\right| $ and $\left[ W_{1/2}^{\pm }(x)\right] \longrightarrow
\left[ W_{-1/2}^{\pm }(x)\right] $.
\bibitem{f8}  Recall that bosonization formulae for the electron
operators are obtained by making a Jordan-Wigner transformation on
the boson operators; likewise the fermionic $(Q,J)$ selection rules ($Q+J$
even) derive directly from the change in statistics: a $(Q,J)$ excitation
of the bosonic LL becomes a $(Q,J^{\prime }=J+Q)$ states due to the
Jordan-Wigner transformation which attaches one flux-tube to each particle ($%
Q$ tubes to $Q$ particles, which increases $J$ by $Q$ units). Since $J$ is
even for the bosonic LL, $Q+J^{\prime }$ is even for the fermionic LL.
\bibitem{f9}  Here and in the remaining of the paper for convenience we will
consider the operators in real space and the wavefunctions will be
localized: the exact eigenfunctions are of course recovered by taking the
Fourier transform of the localized (Wannier) states.
\bibitem{f12}  The Jordan-Wigner transformation which maps spins onto
fermions might lead one to think that the spectrum of the spin chain is
identical to that of the Jordan-Wigner fermions. This is correct only for
the spectrum of an even number of fermions added in (this corresponds
to spin transitions $\Delta S_{z}=2p$ where $p$ is an integer): 
imposing periodic (PBC) or anti-periodic (ABC)  boundary conditions
forces the number of fermions to keep a given parity (even for PBC and odd
for ABC), thus the mapping of spins onto fermions is not valid 
when one adds an odd number of fermions to the ground state. The
spin flip operator is not a Jordan-Wigner fermion since one also creates a
flux tube (the Jordan-Wigner string). This has led to some confusion in the
literature. The $\Delta S_{z}=0$ continuum is correctly
described by the excitations of the Jordan-Wigner fermion system since this
corresponds to the addition of an even number of fermions. 
For the $XX$ spin chain this means that the $\Delta S_{z}=0$
continuum is due to spin one states which are
Jordan-Wigner fermions. But if the mapping were also valid for $\Delta
S_{z}=1$ excitations, this would mean that the continuum is also due to
spin one states. This is not correct because as 
we know this latter  continuum is due to pairs of spinons.
\bibitem{f13}  By adiabaticity, we mean a finite quasiparticles weight, as
for the Landau quasiparticles which is an adiabatic deformation of the
screened free electron.
\bibitem{f14}  The $\widehat{g}$ matrix is distinguished from the $K$ matrix
to avoid a possible confusion because its eigenvalues are $1/K_{c}$ and $%
1/K_{s}$.
\bibitem{holon-charge}  It would be more correct to call this operator an
anti-holon since its electrical charge is $Q_{e}=(-e)Q=-e$; but apart from
the sign of the charge all fractional particles considered in this paper and
their particle-hole conjugates have the same properties. 
So we call this state a holon.
\bibitem{f15}  A naive decoupling of the electron operator would lead to: 
\begin{eqnarray*}
\Psi _{\sigma } &=&\exp -i\sqrt{\pi }\left( \Theta _{\sigma }-\Phi _{\sigma
}\right) \\
&=&\exp -i\sqrt{\pi /2}\left( \Theta _{c}-\Phi _{c}\right) \\
&&\exp -i\sqrt{\pi /2}\left( \Theta _{s}-\Phi _{s}\right)
\end{eqnarray*}
The first term is sometimes identified as the holon and it has a
semionic statistics. The second term is likewise supposed to be the spinon.
These identifications are in general incorrect because these states are unstable
because of chiral separation (see the anisotropic spin chain).
In the special case of $SU(2)$ symmetry, the above
expression for the spinon is accidentally correct.
\bibitem{f16}  The ambiguity which is rather technical is the following:
there are four kinds of anomalous exponents (two chiralities and
two modes). For a given excitation in a given mode, the sum of anomalous
exponents for the two chiralities is fixed but their difference is not. {The
dimensions are therefore only known up to a constant; to fix their values, Frahm and Korepin
make the unproven assumption that they are  perfect squares
if the excitation corresponds to a primary field.}
\bibitem{f17}  Note that in Frahm and Korepin' notations $%
N_{s}=N_{\downarrow }$ is not the spin density.
\bibitem{f18}  The charge matrix and its inverse have the same properties, $%
Z_{2}$ symmetry or $SU(2)$ symmetry in the relevant cases.
\end{references}
\end{document}